\documentclass[aastex]{article}
\usepackage{emulateapj}
\usepackage{psfig}
\def\wisk#1{\ifmmode{#1}\else{$#1$}\fi}
\def\arcpt{\wisk{''\mkern-7.0mu .\mkern1.4mu}}
\def\arcmpt{\wisk{'\mkern-4.0mu .\mkern1.0mu}}

\newcommand{\ha}{H\,$\alpha$}
\newcommand{\hb}{H\,$\beta$}

\newcommand{\oiii}{[O~{\sc iii}]}
\newcommand{\oii}{[O~{\sc ii}]}

\newcommand{\lya}{Ly\,$\alpha$}

\newcommand{\nii}{[N~{\sc ii}]}
\newcommand{\sii}{[S~{\sc ii}]}

\newcommand{\ew}{$W_{\lambda}$}
\newcommand{\etal}{ et al.}

\newcommand{\gl}{$\lambda$}

\newcommand{\kms}{km\,s$^{-1}$}

\newcommand{\msunyr}{M$_{\odot}$yr$^{-1}$}

\newcommand{\degree}{$^{\circ}$}

\slugcomment{To appear in A. J.}

\lefthead{Baker et al..} 
\righthead{Tunable filter imaging of quasar field MRC\,B0450--221 }

\begin{document}

\title{Tunable-filter imaging of quasar fields at $z\sim 1$.
I. A cluster around MRC\,B0450--221}

\author{Joanne C. Baker\altaffilmark{1,2,3},
Richard W. Hunstead\altaffilmark{4},
Malcolm N. Bremer\altaffilmark{5,6,7},
Joss Bland-Hawthorn\altaffilmark{8},
Ramana M. Athreya\altaffilmark{6,9},
Jordi Barr\altaffilmark{5}}

\authoremail{jcb@astro.berkeley.edu}

\altaffiltext{1}{Hubble Fellow. }
\altaffiltext{2}{Astronomy Department, 601 Campbell Hall, 
	University of California, Berkeley, CA 94720, USA.}
\altaffiltext{3}{Astrophysics, Cavendish Laboratory, Madingley Road, 
        Cambridge, CB3 0HE, UK.}
\altaffiltext{4}{School of Physics, 
       University of Sydney, NSW 2006, Australia. }
\altaffiltext{5}{Department of Physics, Bristol University, 
	H.H. Wills Laboratory, Tyndall Avenue, BS8 1TL, UK.}
\altaffiltext{6}{Institut d'Astrophysique de Paris, 98bis Boulevard Arago, 
	75014 Paris, France.}
\altaffiltext{7}{Sterrewacht Leiden, P.O. Box 9513, 2300RA Leiden, 
	The Netherlands.}
\altaffiltext{8}{Anglo-Australian Observatory, P.O. Box 296, 
	Epping, NSW 1710, Australia.}
\altaffiltext{9}{European Southern Observatory, Casilla 19001, 
	Santiago 19, Chile.}

%************
%  ABSTRACT *
%************

\begin{abstract}

Using a combination of multicolour broad- and narrow-band imaging 
techniques and follow-up spectroscopy, we have detected an overdensity of
galaxies in the field of quasar MRC\,B0450$-$221, whose properties are 
consistent with a cluster at the quasar redshift $z=0.9$. 
An excess of red galaxies ($V-I>2.2$, $I-K'\geq 3.8$) is evident  
within $1'$ of the quasar, with the colours expected for galaxies 
at $z=0.9$ that have evolved passively for 3\,Gyr or more. 
A number of line-emitting galaxies (nine candidates with 
equivalent widths \ew$>70$\,\AA) are also detected in the 
field using the TAURUS Tunable Filter (TTF). Three have been 
confirmed spectroscopically to indeed lie at $z=0.9$.
The TTF candidates with the strongest \oii\ line emission cluster in a group
which lies 200--700\,kpc away from the quasar and the red galaxy excess,
and therefore most likely on the outskirts of the cluster. 
These observations are the 
first in a series probing quasar environments at $z\sim 1$ with TTF. 

\end{abstract}


%**************
% INTRO       *
%**************

\section{Introduction}
\label{sec:intro}

Clusters and groups of galaxies are notoriously difficult to find 
at redshifts $z\sim 1$ and beyond. The two conventional methods 
--- X-ray and blind optical surveys ---  
are  difficult and highly incomplete beyond about $z= 0.3$.
Optical surveys are hampered by the lack of contrast
between the distant cluster and foreground galaxies, and 
X-ray surveys are biased towards detecting  the most massive 
(and/or merging) clusters at high redshift. 
An efficient way to initiate a search for distant galaxy clusters is to 
use markers already known to be at high redshift, such as active galactic 
nuclei (AGN). Radio-loud AGN are favoured as pointers to high-$z$ structures
because they can be seen out to the highest redshifts and 
previous studies have found they tend to inhabit
environments of above average galaxy density
at least out to $z\sim 0.7$ (Yee \& Green 1987).
In addition, selection of radio sources avoids the inadvertent omission
of objects which are highly obscured by dust. 
AGN environments may also play a key r\^ole in the luminosity evolution of
AGN --- at moderate redshifts high-luminosity AGN are found in rich 
clusters, whereas at $z\lesssim 0.3$ only low-luminosity AGN reside in the
richest clusters (Ellingson, Yee \& Green 1991; Hill \& Lilly 1991).

Most galaxy clusters detected around AGN have been identified on the
basis of statistical overdensities of objects found
within some radius of the AGN (e.g. Yee \& Green 1987; Hutchings 1995;
Hall, Green \& Cohen 1998). 
However, a better strategy to overcome foreground dilution 
is to identify individual high-redshift galaxies by, for example, 
their characteristic colours or emission 
lines. Perhaps easiest to distinguish are the very red galaxies 
characteristic of passively evolving ellipticals found in cluster 
cores (Smail et al. 1998). However, these red galaxies may be 
few in number as the fraction of blue galaxies in clusters is 
known to increase substantially at $z\gtrsim 0.3$ (Butcher \& 
Oemler 1984; Dressler \& Gunn 1992), possibly reaching proportions as 
high as 80\% of the cluster population at $z\approx 0.9$ 
(Rakos \& Schombert 1995). Therefore, the search for blue, 
star-forming galaxies must form a crucial part of any 
survey for distant clusters. 
However, there have been very few observations of emission-line
galaxies around AGN, mainly as a result of poor sensitivity 
due to the relatively wide pass-bands
and low efficiency of monolithic narrow-band filters in the 
optical and infrared. 
In addition, there have been no systematic surveys of complete
samples of AGN, as this would require obtaining a large set of
filters to cover a range of redshifted pass-bands. 
Nevertheless,  groups of emission-line galaxies have indeed 
been noted in a number of quasar fields at $z\sim 1$--2 
(Hutchings, Crampton \& Persram 1993; Hutchings 1995). 

\begin{figure*}
%%\centerline{\psfig{file=four.eps,bbllx=
\centerline{\psfig{file=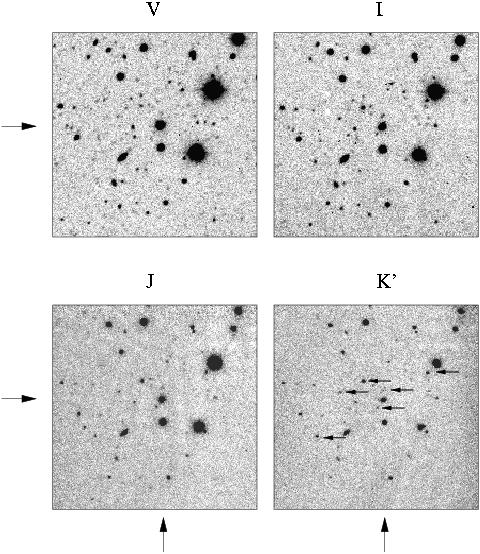,bbllx=
66pt,bblly=117pt,bburx=544pt,bbury=673pt,height=17cm} }
\caption[e]{\footnotesize
Images of the central $190'' \times 190''$ of the MRC\,B0450--221 field
in $V, I, J, K'$ (i.e. to match the infrared field size). 
The quasar is identified by the external arrows. N is up, E is left. 
Note a number of red galaxies visible in $I$ (and $J, K'$) but not $V$, 
including the two galaxies nearest the quasar. The most prominent 
examples are marked by arrows on the $K'$ image. 
}
\label{four}
\end{figure*} 
\newpage

From a broader perspective, both 
field and cluster samples (Ellis et al. 1996; Cowie et al. 1996, 1997; 
Hammer et al. 1997; Balogh et al. 1997, 1998)
indicate that more star-formation is
occurring on average at $z \sim 0.5$--1 than in the nearby universe
(e.g. Butcher \& Oemler 1984; 
Lilly et al. 1996; Madau et al. 1996; Connolly et al. 1997;
Madau, Pozzetti \& Dickinson 1998 but see Cowie, Songaila \& Barger 1999). 
In addition, star-formation is apparently reduced in the centres of massive 
clusters (Balogh et al. 1997, 1998), suggesting star-formation is affected
by environment, although the processes causing this are unclear. 
Both spectroscopic field and cluster samples, however, 
become incomplete at about $z\sim 1$. At this 
redshift and beyond, narrow-band imaging 
techniques are the most efficient in detecting emission-line galaxies. 
In fact galaxies out to the highest redshifts 
have been found in this way by their strong \lya\ emission 
(Hu, Cowie \& McMahon 1998; Hu, McMahon \& Cowie 1999).
Narrow-band imaging 
has the obvious advantage that galaxy selection is made directly 
on the basis of high star-formation rate, so the most active
systems are found most easily. 

Another reason for pursuing emission-line galaxies in distant 
clusters and quasar environments is that blue galaxies tend to dominate
the outer regions of rich clusters, where
they are sensitive probes of cluster sub-structure. 
The study of sub-structure is very important at high redshifts, where the 
aggregation of clusters from smaller groups is expected 
if clusters formed hierarchically (e.g. Jenkins \etal\ 1998). 

 A powerful way, therefore,  of searching for galaxies around 
high-redshift AGN is the combination of multicolour imaging 
(to detect the red population) 
and narrow-band imaging (to detect emission-line galaxies; ELGs). 
Using both these methods, we have set out to target systematically 
the fields of a  representative sample of radio-loud 
quasars (drawn from the Molonglo Quasar Sample; see Kapahi \etal\ 1998). 
Full results for the survey will be published elsewhere;
preliminary results are given in Bremer \& Baker (1998) and 
Bremer, Baker \& Lehnert (2000). 
This study benefits in particular from deep narrow-band imaging carried out 
using the TAURUS Tunable Filter (TTF) --- a 
Fabry-Perot-based imaging system developed at 
the  Anglo-Australian Telescope (AAT). TTF enables
sensitive narrow-band imaging with spectral resolution ${\cal R} \approx
100$--1000, 
tunable over the window 3700--10000\,\AA\ (Bland-Hawthorn \& Jones 1997, 
1998). With the high throughput and narrow bands attainable with TTF
we achieve high sensitivities
(e.g. $10^{-17}$ erg\,cm$^{-2}$s$^{-1}$arcsec$^{-2}$ ($3\sigma$)
in 12 minutes of integration),  which are sufficient to 
probe \oii\ emission in galaxies with star-formation rates
greater than a few \msunyr\ at $z\sim 1$. 
TTF data for the full sample will be presented in a future paper.

In this paper, we report the detection of a cluster
in the first field studied, MRC\,B0450--221, a quasar at $z=0.898$.
This is inferred from an excess of red galaxies near the quasar 
detected in broad-band images, and also an excess of emission-line galaxies
detected in the field. 
The observations (TTF, broad-band imaging and follow-up spectroscopy) 
and analysis are described in detail in 
Section \ref{sec:obs}, and results presented in 
Sections \ref{sec:ttfres} and \ref{sec:optir}. 
These include the detection, and confirmation with 
multi-object spectroscopy, of \oii-emitting
galaxies at  $z=0.9$ in the field of MRC\,B0450--221 with TTF.
Cosmological parameters of $H_0 = 50$ \kms\,Mpc$^{-1}$, 
$\Omega = 1$ and $\Lambda = 0$ are taken throughout.

\section{Observations}
\label{sec:obs}

\subsection{Broad-band optical and infrared imaging}
\label{sec:optir}

Broad-band optical imaging of the field of 
MRC\,B0450--221 was carried out using
EFOSC-I at the ESO 3.6m telescope at La Silla 
on 1997 March 12. Deep images in $V$ (1800s) and Gunn 
$i$ (3600s) filters were obtained in average seeing of 1\arcpt 2.
The field of the $512 \times 512$ pixel Tektronix CCD
covered 5\arcmpt 2 $\times$ 5\arcmpt 2 with 0\arcpt 61 pixels. 
Photometric standard stars and twilight-sky flat fields were 
observed in all filters. The data were bias-subtracted,
flat-fielded and photometrically analysed using standard 
methods in IRAF and FOCAS (Valdes 1993). Limiting 
magnitudes ($2.5\sigma$ 
in a three-pixel radius aperture) are $V=24.8$ and $I=23.1$. 
The $I$-band photometry was carried out within the 
standard Bessell/Cousins system (colour term corrections
from using the Gunn $i$ filter are less than 0.1\,mag and so
have been neglected). Galactic extinction corrections have not 
been applied as the extinction is low along this sightline
($E_{\rm B-V}=0.04$, $A_{\rm V}=0.14$; Schlegel et al. 1998).

Infrared images were obtained on 1996 February 8--9 with the
ESO 2.2m telescope using the IRAC-2B camera equipped with a 
$256 \times 256$
HgCdTe array (0\arcpt 506 per pixel). A field of diameter 2\arcmpt 5 
centred on the quasar was observed in the $J$ and $K'$ bands
for a total of 1 hour each. The total integration time was 
subdivided into many short-exposure frames (30 in $J$, 60 in 
$K'$) which were dithered in a non-repeating pattern on the sky. 
The images were dark-subtracted, flat-fielded and the sky was 
removed in a standard fashion using specially written IDL routines. 
Photometric calibration was achieved by observing UKIRT faint 
standard stars FS11, FS14 and FS27.

Infrared object detection and photometry were carried out using 
FOCAS (Valdes 1993) over a region $118'' \times 136''$
around the quasar, where the noise in the mosaic image was lowest.
The infrared $J$ and $K'$ images are shown in Figure \ref{four} 
together with the optical $V$ and $I$ images for the same area.
In order to detect all objects visible 
by eye in the infrared fields, the detection threshhold was set
to 1.5 times the background noise level of the combined frame 
for objects comprising 4 contiguous pixels or more. 
Magnitudes (measured within 3\arcpt 0-diameter circular apertures) 
were determined for all 44 objects in the field, of which
33 were fainter than $K' = 17.5$ and detected in both the $J$ and $K'$ frames. 
Limiting magnitudes ($1.5\sigma$) are $K'= 20.5$ and $J = 21.1$
for the mosaic images.
Photometry for these 33 objects is listed in Table \ref{irtab}.
Of these faint objects, ten were at or below the detection limit 
in $V$, and two were detected  neither in $I$ nor $V$
in the optical images (to approximate limits $V$=25, $I$=23.5).
The colours of the two reddest  objects, $I-K'>4$, $J-K'=2.7$ and 2.1
and $K'\approx 19$, are consistent with very red galaxies seen in 
other fields, including so-called extremely red objects (EROs;
Elston, Rieke \& Rieke 1988; Thompson et al. 1999).

\subsection{Tunable filter imaging: observations}
\label{sec:ttfsetup}

\begin{figure*}
\centerline{\psfig{file=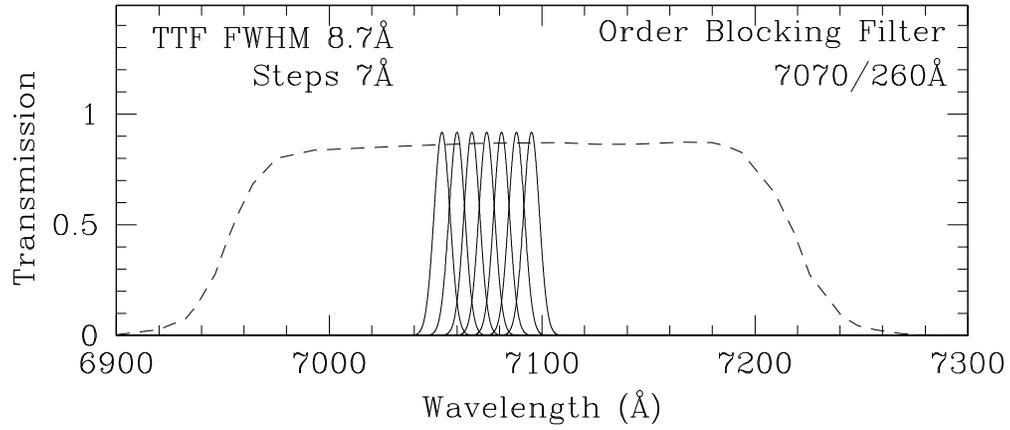,bbllx=
22pt,bblly=165pt,bburx=588pt,bbury=420pt,height=6cm} }
\caption[fig2]{\footnotesize
Passband sampling used in the TTF observations (profiles 
approximated by Gaussians; solid lines). The wavelengths shown 
are as measured at the location of the 
quasar near the field center (note the wavelength sampled 
varies radially across the field for a given observation, see text). 
The transmission profile of the 7070/260\AA\ order-blocking filter
is shown as a dashed line. 
}
\label{pass}
\end{figure*}

The quasar MRC\,B0450$-$221 was observed with TTF on the TAURUS-2
Fabry-Perot Imager at f/8 on the AAT on 1997 February 3.  The 
Tek 1024$\times1024$ pixel CCD detector was used, with a resulting pixel scale 
of 0\arcpt 60 over a $10'$-diameter circular field. TTF was tuned 
at the field centre to give an observed bandpass of 8.7\AA\ FWHM centred on 
the wavelength of \oii\,\gl3727 
at the quasar redshift, $z=0.898$.
The efficiency of the narrow observing band is high, being
dictated only by the high throughput of the wider 7070/260\AA\ 
order-sorting filter (about 90\% peak response) and the 
instrumental throughput of TTF (again about $90$\%). The 
transmission profiles of the order-blocking filter and the
the individual scans are shown in Figure \ref{pass}.

A wavelength scale for the TTF observations
(mapping instrumental units, $Z$, to wavelength, 
$\lambda$, for different etalon spacings) was calibrated 
at the field centre
(and also at the eastern edge of the field) to within 1\AA\ by 
scanning through lines in a CuAr comparison lamp spectrum.
Example settings and parameters for a typical
TTF setup used are given in Table \ref{tabset}. 
Because the transmitted wavelength 
varies due to the differing angles of incidence 
of light rays entering the Fabry-Perot etalon, there is 
a radial dependence of transmitted wavelength away from 
the position of the optical axis. Thus arc or night-sky OH lines 
appear as rings or arcs. To ease removal of cosmic rays and ghost images, 
the etalon was tilted so that the optic axis lay 280 pixels to 
the south of the field centre. To illustrate the off-centre geometry
and show the amount of variation in transmitted wavelength across the field, 
Figure \ref{wavecal} shows the TTF image with wavelength 
contours superimposed at intervals of 7\,\AA. Thus, for a single etalon 
spacing, there is a north-south gradient in transmitted wavelength of about 
35\AA\ in $7'$, and approximately 
14\AA\ difference between the centre and the east/west edges of 
the field.

\begin{figure*}
\centerline{\psfig{file=baker.fig3.ps,bbllx=
104pt,bblly=194pt,bburx=509pt,bbury=595pt,height=9cm} }
\caption[fig3]{\footnotesize
Illustration of wavelength variation across the TTF field. 
Contours are drawn at 7\AA\ intervals, on top of the 
$7' \times 7'$ TTF image (shown smoothed and at high 
contrast).  The
centre of the ring pattern (longest wavelength) is marked ($+$).
Wavelength decreases radially outward from this point.
The quasar is marked between bars. N is up, E left.
}
\label{wavecal}
\end{figure*} 

TTF scanned the quasar field in seven consecutive steps of 7\AA\ 
(frames F1 to F7) either side of and 
centred on \oii\ at the quasar redshift. At this redshift, 
the steps correspond to velocity intervals of 300\,\kms. 
Total exposure times were 1000s 
in each band, and the seeing typically 1\arcpt 5 FWHM. Multiple exposures 
were taken, dithered by about $10''$ on the sky, to allow the 
unambiguous rejection of cosmic rays and ghost images.
A spectrophotometric standard star (LTT\,2415) was observed at 
the central TTF wavelength setting of the scan. 
The TTF photometry is expressed thoughout in 
AB magnitudes (denoted $I(AB)$ here). 
Dome flats were also taken through TTF at each wavelength setting. 

\subsection{Tunable filter imaging: data reduction}

The raw TTF images were bias-subtracted and flat-fielded using standard 
procedures in {\sc iraf}. Emission from night-sky lines, visible 
as rings in the Fabry-Perot images, was removed using median frames 
to within two percent of the background level. This reduction
produced a set of seven images, one for each of the narrow bands. 
Limiting magnitudes were $I(AB)=21.4$ in each individual frame 
and $I(AB)=22.4$ for a summed image of all 7 frames 
($2.5\sigma$ in a 3-pixel radius circular aperture).

\begin{figure*}
%%\centerline{\psfig{file=strips.eps,bbllx=
\centerline{\psfig{file=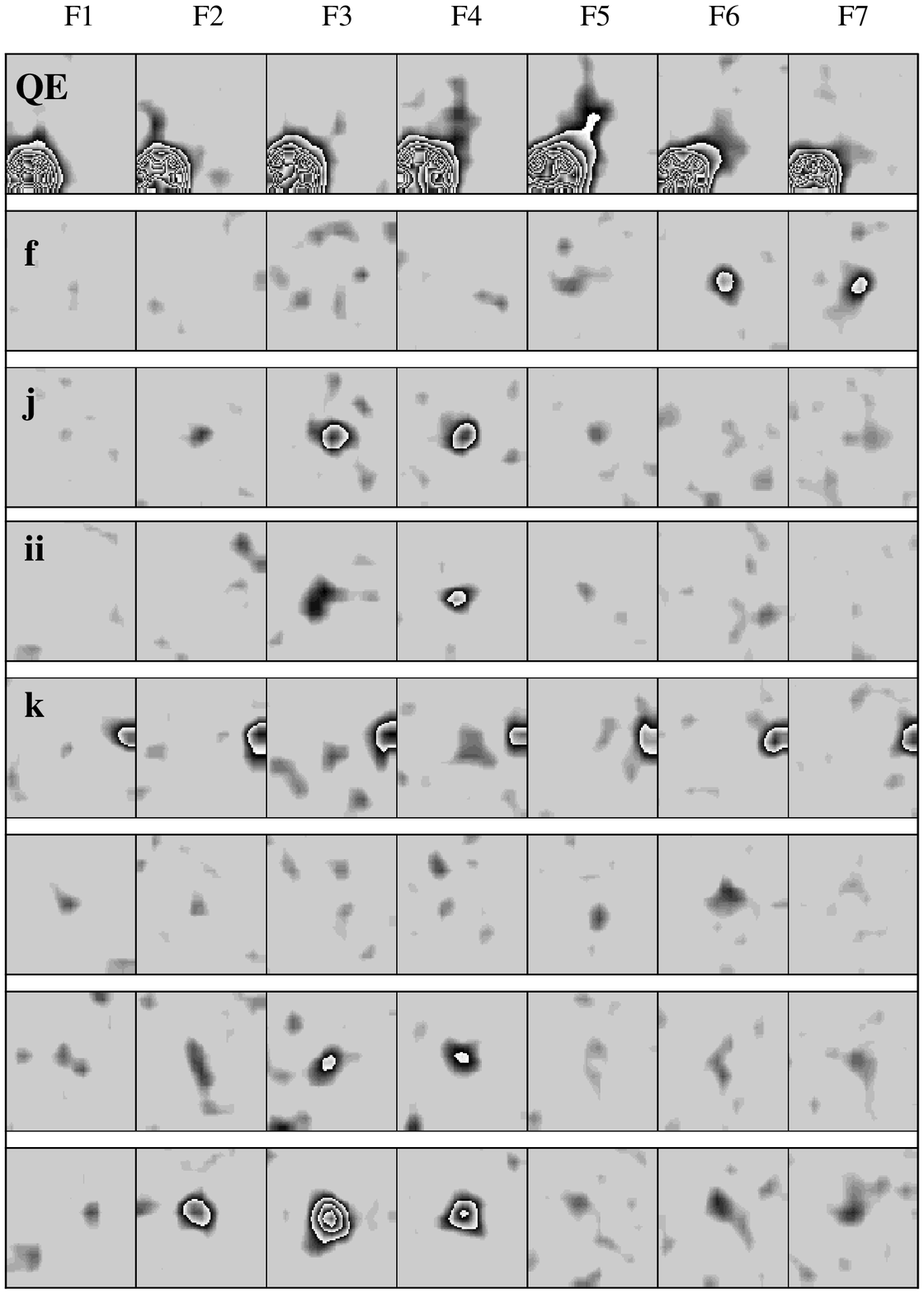,bbllx=
83pt,bblly=83pt,bburx=536pt,bbury=715pt,clip=,height=12cm}}
\caption[fig4]{\footnotesize
A montage of TTF images showing examples of candidate line-emitters
in the field of MRC\,B0450$-$221.
Each row corresponds to an object, and 
shows images ($12''$ square) drawn
from the sequence of TTF frames (F1 to F7),
i.e. spanning a total wavelength range of 49\,\AA,
in seven adjoining passbands near the quasar redshift (see text).
Those which have been targetted spectroscopically 
have been labelled (as in Figure \ref{mosids}), 
as well as the quasar extended emission (QE).
All images have been smoothed heavily for clarity
and are shown on the same linear (wrapping) greyscale.
The top panel clearly shows \oii\ emission immediately to the northwest of
the quasar (bright object in lower left corner of images). 
The images are oriented so that N is up, E to the left.
}
\label{ttf}
\end{figure*}

\begin{figure*}
\centerline{\psfig{file=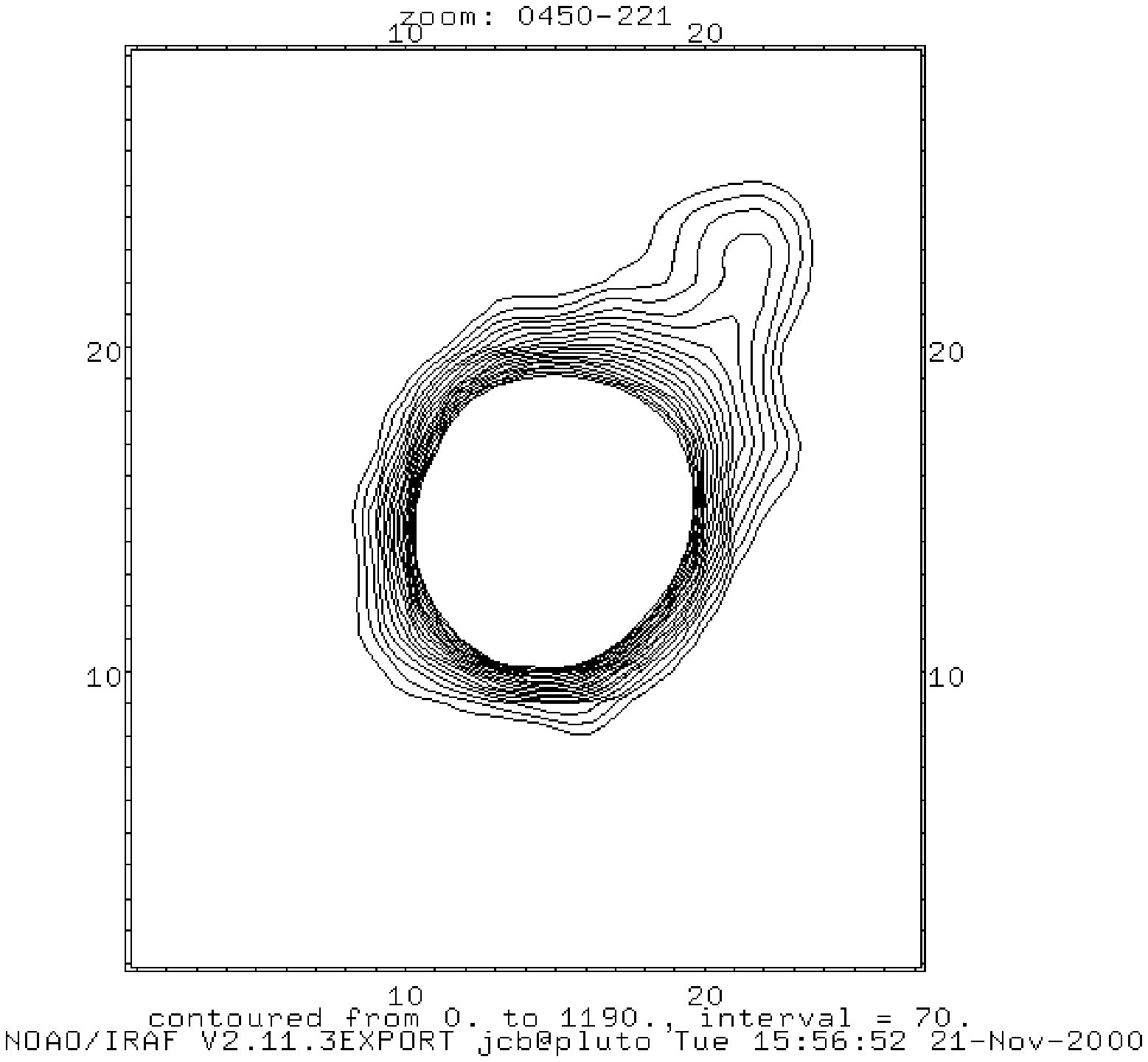,bbllx=
120pt,bblly=214pt,bburx=475pt,bbury=599pt,clip=,height=6cm}}
\caption[fig5]{\footnotesize
Contour plot of co-added TTF data for the quasar, showing an extension
to the northwest. The axes are labelled
in pixels (0\arcpt 60). 
The contours run from 0 to 1200 counts in steps of 70. 
N is up, E is left.
}
\label{qcontour}
\end{figure*}

\begin{figure*}
\centerline{\psfig{file=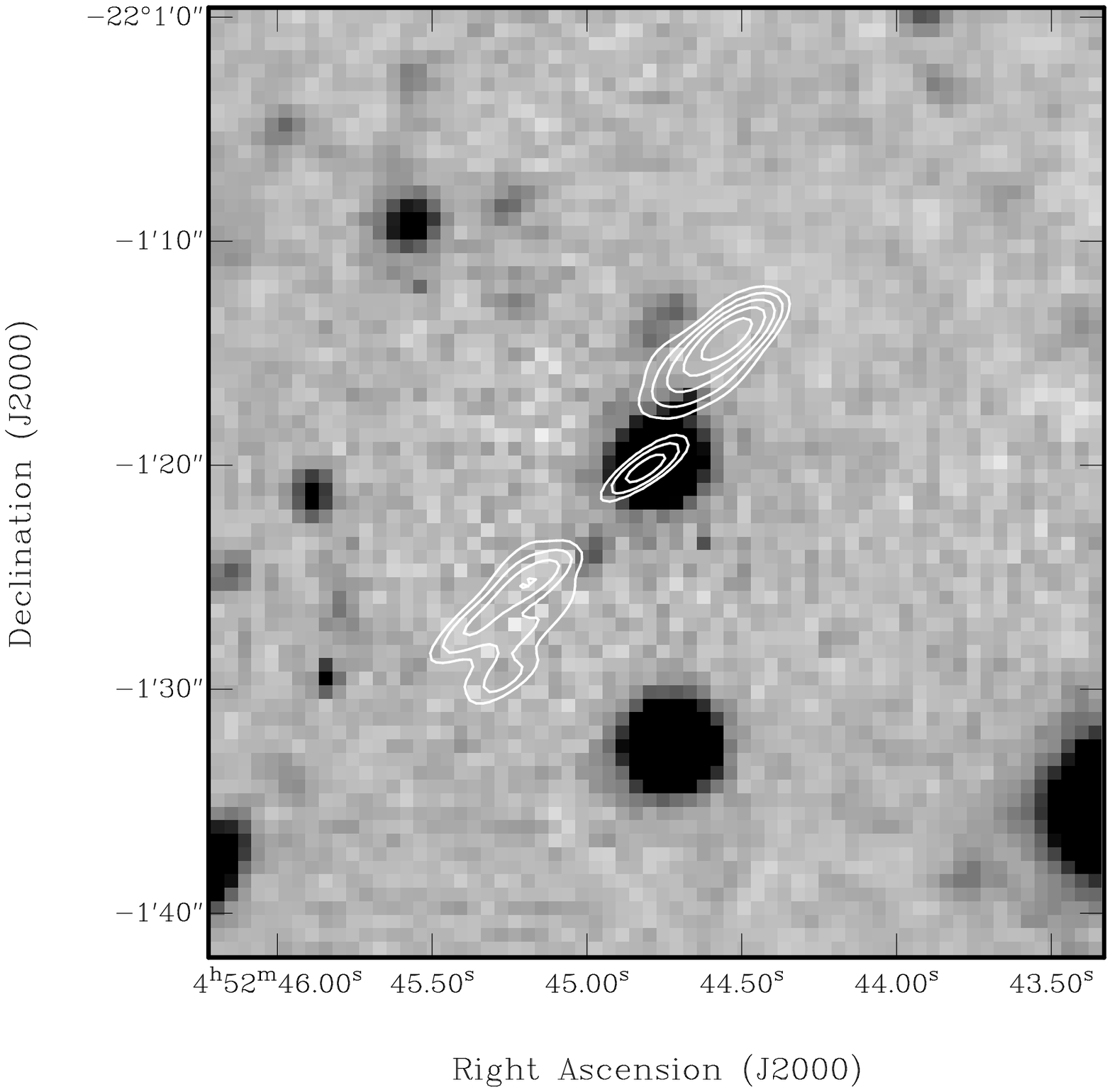,bbllx=
55pt,bblly=143pt,bburx=559pt,bbury=631pt,clip=,height=8.5cm}}
\caption[fig6]{\footnotesize
Overlay of radio contours with $I$-band image for the quasar.
The radio data are from VLA snapshot observations at 4.8\,GHz, as
described in Kapahi et al. (1998). The contour levels are
4\,mJy/beam$\times(0.6,1.2,2.4,4.8,9.6)$, and the peak flux 
density is 60.8\,mJy/beam. The major and minor axes of the restoring
beam are 2\arcpt 8 by 0\arcpt 9, and its orientation is PA=-57$^{\circ}$.
The unresolved radio core itself is difficult 
to see against the quasar on this reverse greyscale image, 
but it aligns well with the optical counterpart.   
}
\label{radio}
\end{figure*}

Objects in the field which brightened in one or two bands were clearly 
visible by eye in the TTF images --- examples are shown in 
Figure \ref{ttf}, including the quasar itself (bright object
offset from centre in the top panel). Extended emission
is clearly visible 
to one side of the quasar peaking about 300\,\kms\ redward of the
published redshift (sampled in F4), and clearly visible over 
the range 0--1000\,\kms. Figure \ref{qcontour} shows a contour image
of the extended emission-line region.
The close alignment of this emission with the radio jet axis 
(Kapahi et al. 1998; Figure \ref{radio}) 
suggests it is physically associated with the
quasar outflow, but we note the possibility that it may be 
instead a separate galaxy projected nearby. 
If the extended emission is associated with the quasar,  
then its slightly higher redshift may indicate that
the true systemic redshift of the quasar is closer to
$z=0.900$ rather than $z=0.898$. Extended narrow-line emission is 
generally taken to be a better indicator of the systemic redshift 
than the broad emission lines, which may show systematic velocity shifts 
relative to the narrow lines of order several hundred \kms\ 
(Gaskell 1982; Tytler \& Fan 1992). However, the
published redshift, $z=0.898$,
was based essentially on broad Mg\,{\sc ii}$\lambda2800$ 
(Hunstead, Murdoch \& Shobbrook 1978) which is rarely shifted 
far from the systemic redshift on average
(Tytler \& Fan 1992).
For consistency we will continue to use the published 
redshift ($z=0.898$) as a reference point throughout this 
paper\footnote{A recent spectrum of quasar MRC\,B0450--221,
taken after this paper was submitted, indeed 
shows a more accurate redshift of $z=0.900$ 
(de Silva et al., in preparation).}. 

As shown in Figure \ref{radio} and described by Kapahi et al. (1998),
the radio morphology of MRC\,B0450--221 is that of a classic triple
FR\,II source. It is lobe-dominated, with  a steep spectrum and 
radio core-to-lobe luminosity ratio (at 10\,GHz, restframe) $R_{10}=0.086$.
The source has a total projected extent of $18''$ oriented 
in position angle PA=135\degree. Given the limited dynamic range and
resolution of the 
radio image, there are no obvious indications of strong interactions 
between the radio plasma and any intracluster medium (e.g. tails, 
bent structure). However, the source does follow the correlation of
closer (northern) radio lobe falling on the same side as the extended optical 
emission (McCarthy, van Breugel \& Kapahi 1991), which might indicate 
higher ambient density on this side.
Further study of the radio-optical
alignment is deferred until better quality radio data are available.

\subsection{TTF photometry and object detection} 

\begin{figure*}
%%\centerline{\psfig{file=complete.ps,bbllx=
\centerline{\psfig{file=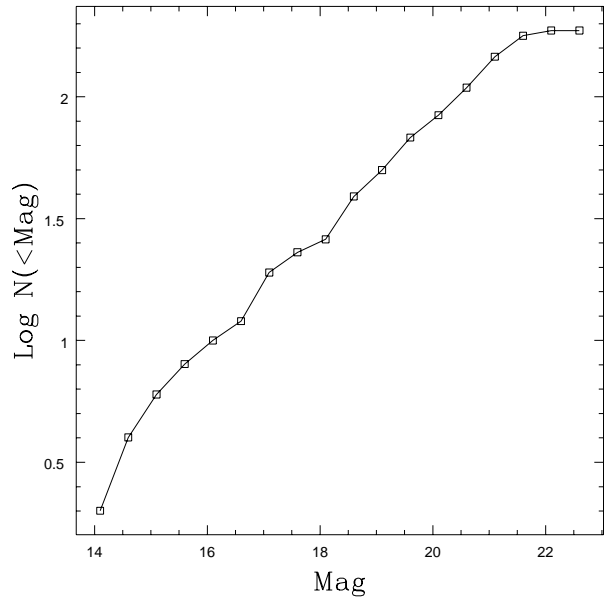,bbllx=
22pt,bblly=148pt,bburx=568pt,bbury=698pt,clip=,height=8cm}}
\caption[fig7]{\footnotesize
Number of objects extracted in the $7' \times 7'$ frame 
brighter than a given TTF (I(AB)) magnitude. This shows the counts are 
complete to $I(AB)\approx 21.5$. 
}
\label{complete}
\end{figure*}

Object detection and aperture photometry were carried out 
on the TTF frames using FOCAS (Valdes 1993). To avoid edge effects, 
the circular TTF images were first trimmed down to the 
central $7' \times 7'$ square. The images were summed  
to form a deep image for object detection purposes. 
Objects were identified in the
co-added image using a three-pixel aperture radius (1\arcpt 8) 
and $2.5\sigma$ detection threshhold, to optimise sensitivity 
to objects near the detection limit. 
Aperture photometry was
then carried out using the same aperture template for each
TTF frame --- fixed apertures (rather than isophotes)
were preferred in order that the
noise uncertainties be well defined. Because small apertures were used, 
a constant 0.4\,mag offset was applied to each object (at the end of 
the analysis described in Section \ref{sec:elg}) to 
correct for missing flux.
This offset was measured by comparing magnitudes measured with 
the three-pixel apertures and larger apertures (e.g. five and seven-pixel 
radii) for both stars and galaxies over the whole field.
The offset was constant over the whole magnitude 
range $I(AB)=14-22$ within the photometric uncertainties. 
By matching the catalogues generated in this way, a 
preliminary database was created containing
a set of magnitudes in each TTF frame 
for each object in the field. 
As shown in Figure \ref{complete}, 
object detection by this process was essentially complete to $I(AB)=21.5$.

\subsection{Multi-object spectroscopy}

Follow-up multi-slit spectroscopy was carried out on 14 objects 
in the field of MRC\,B0450--221 using EFOSC-2 on the ESO 3.6-m 
telescope at La Silla on 1998 September 28.
A 300-line/mm grating was used with the Loral CCD detector 
(with two-pixel binning in both the spatial and spectral directions)
to attain 10\AA\ spectral resolution over the approximate wavelength 
range 5500--10000\AA. Thirteen slitlets, oriented
east-west, were used to target 
simultaneously two bright stars and 12 faint galaxies
(typically $I(AB)=21$--22) near the quasar. 
The targets included five convincing ELG candidates detected in 
the TTF analysis (described in Section \ref{sec:elg}), 
a red galaxy, and  other targets chosen simply in order to fit as
many candidates as possible near to the quasar in the slit mask
(see Section \ref{sec:mos} for specific targets and results). 
The field was observed for a total of four hours,
split into 30-min exposures. Due to time constraints
(mainly due to weather) we were only able to observe with a single 
slit mask. Seeing was variable $1''-2''$, and
conditions were not photometric. 

The individual images were 
first bias-subtracted and flat-fielded using 
standard procedures in IRAF. Then the images were co-added and 
cosmic rays removed. For each slit, the background
emission from the night sky was interpolated and subtracted, 
and the final spectrum extracted. Wavelength calibrations were 
carried out using observations
of a CuAr comparison lamp taken through the slit mask.

\section{Identification of ELGs in TTF data}
\label{sec:ttfres}

\subsection{Line identification method}
\label{sec:elg}

In order to distinguish emission-line galaxies peaking in 
any one of the seven TTF frames, an automated method was adopted.
First, photometric zeropoint offsets between frames 
were corrected by comparing the averages of 
the measured magnitudes of bright unsaturated stellar 
objects ($I(AB) =15$--17) in the field. The final corrections
were less than 0.2\,mag in all the frames, and were made 
relative to the central-wavelength  frame (F4) --- this effectively 
removed  systematic offsets between frames to within 0.02 magnitudes. 
Next, for each object in the field, the frame in which the raw
counts for that object were a maximum, i.e. the peak flux, was determined. 
The significance of this peak was calculated relative to 
the average counts measured in the other frames, after 
excluding the one in which the peak occurred and the (one or) two 
frames adjacent to it. 
The neighbouring frames were excluded because the
emission lines typically spread flux over two or more frames. Objects
where the difference between peak (`line') and average off-peak
(`continuum') counts exceeded 2.5 times the rms sky noise in that bin 
($\sigma$) were considered to be preliminary emission-line candidates. 
This level 
(rather than a more conservative $5\sigma$) was adopted as a first cut
in order to include as many real detections 
as possible, given the relatively short integration times for 
these preliminary observations and that line flux may be spread over
two or more bands.  A wavelength for the peak 
was assigned to each object according to the spatial dependence 
of wavelength discussed in Section \ref{sec:ttfsetup}.

To exclude faint M-stars --- which have a strong TiO strong 
absorption feature at 7050\AA\ and which might otherwise contaminate 
the data over our small wavelength window ---  
we only considered candidates whose peak fluxes
fall at wavelengths longer than 7050\AA. 
According to Martini \& Osmer (1998) in a comparable deep multicolour 
survey, we would expect of order 35 M-stars in a field of 
$7' \times 7'$ down to $V=22.5$, which is of order the 
number excluded by this cutoff criterion (i.e. 53 objects, which
includes some whose signal is primarily noise and so excluded at random).

\begin{figure*}
%%\centerline{\psfig{file=magdiffgt7050.ps,bbllx=
\centerline{\psfig{file=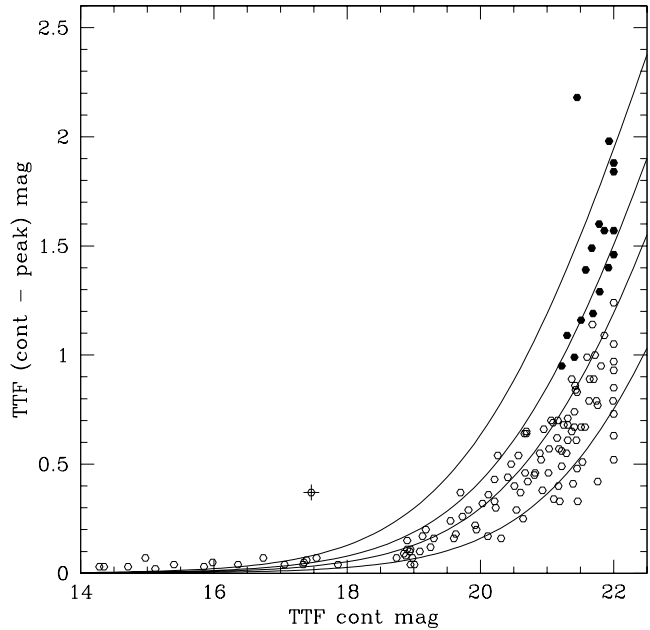,bbllx=
22pt,bblly=156pt,bburx=573pt,bbury=700pt,height=8.5cm} }
\caption[fig8]{\footnotesize
Difference between peak and average continuum magnitudes 
measured in all TTF frames (i.e. `line strength') 
plotted as a function of estimated continuum
magnitude for all objects detected in the co-added TTF image.
The plotted curves show 1, 2, 3 and 5 times the rms 
uncertainties ($\sigma$) due to noise. Objects with 
$I(AB)>21$ and peaks exceeding the continuum by amounts greater than 
$2.5\sigma$ have filled symbols. The quasar is marked by a `+'. 
The two objects just above the 3-$\sigma$ line between 20 and 21 mags
appear to be spiral galaxies, probably at lower redshift than the
quasar system.
}
\label{offpeak}
\end{figure*}

The difference between the `continuum' and peak magnitudes 
is plotted in Figure \ref{offpeak} as a function of
continuum magnitude for all objects detected in the field
(and peaking longer than 7050\AA).
Curves are shown corresponding to 1, 2, 3 and 5 times the rms noise
errors ($\sigma$).
The lower envelope on Figure \ref{offpeak} is close to the 
$1\sigma$ curve as expected; in seven bins, essentially all objects 
should show at least a $1\sigma$ peak due to noise statistics alone. 
Clearly, more faint objects are detected
with peaks above $2.5\sigma$ than expected due to noise alone; 
for objects fainter than  $I(AB)=21$, 17/68 have peak fluxes 
falling  above the $2.5\sigma$ level, 
of which five lie above $4\sigma$ in Figure \ref{offpeak}.
The quasar itself lies well above the noise curves.

We now consider ELG candidates with the following 
properties: (i) excess emission greater than 2.5 times the rms 
sky noise in at least one band, and (ii) average continuum magnitude
fainter than $I(AB)=21.0$. As mentioned above, 
17 candidates fulfill these criteria.
This magnitude cutoff was chosen to
select the most massive ellipticals at $z=0.9$ and fainter galaxies. 
The brightest member of 
a typical cluster at $z=0.9$ is expected to have $I(AB)\approx 20$--21
(Hoessel, Gunn \& Thuan 1980), so several tens of cluster galaxies 
would be expected to be seen down to $I(AB)\approx 23$. This is also 
consistent with the magnitude-$z$ relation for 3CR radio galaxy hosts,
which are typically $I(AB)\approx 21$ at $z=1$ 
(Eales 1985; Snellen et al. 1996)
and may be progenitors of central cluster galaxies 
(Best, Longair \& R\"ottgering 1998). We note that a brighter limit, 
$I(AB)=20.0$, would only have introduced two extra candidates. 
Later, we will consider only ELG candidates with high 
equivalent widths as the strongest candidates.

\subsection{Velocity distribution of ELG candidates}

\begin{figure*}
\centerline{\psfig{file=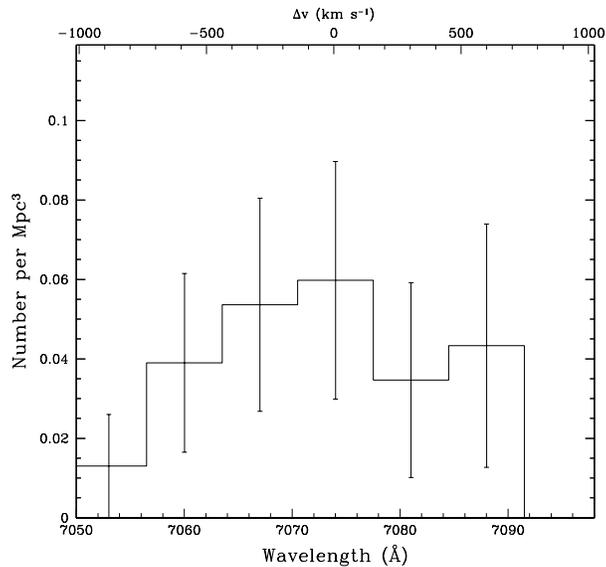,bbllx=
18pt,bblly=145pt,bburx=585pt,bbury=701pt,height=8cm}}
\figcaption[fig9]{\footnotesize
Space density of ELG candidates detected with TTF 
as a function of wavelength (lower axis)
and velocity (upper axis) relative to the quasar redshift $z=0.898$. 
Bins of 300\,\kms\ were used (i.e. 7\AA).
For comparison (see Section \ref{sec:disc}),
the expected space density of field galaxies with similar
star-formation properties is about 0.001\,Mpc$^{-3}$ at $z=0.9$
(Cowie et al. 1997). Poisson error bars are shown. 
}
\label{nompc}
\end{figure*}

\begin{figure*}
\centerline{\psfig{file=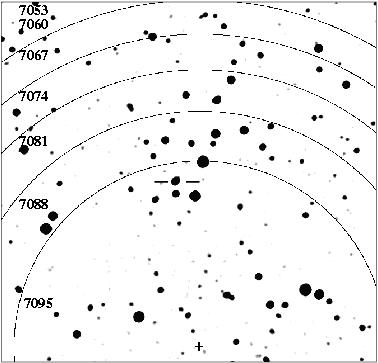,bbllx=
122pt,bblly=212pt,bburx=495pt,bbury=575pt,height=8cm}}
\caption[fig10]{\footnotesize
Northern limits of wavelength coverage for 7\AA\ 
slices used in Figure \ref{nompc}, displayed on a high-contrast
coadded image of the TTF field.
The central wavelength (\AA) of each slice is labelled just
below each arc; 
all the area from the arc to the southern edge of the field
is sampled in that wavelength range. The 7053\AA\ and 7060\AA\
bins sample the entire field. 
The optic axis is marked with a plus, and the quasar identified 
between horizontal bars.
N is up, E is left.
}
\label{waveimage}
\end{figure*}

To examine the distribution of ELG candidates as a function of wavelength,
the space density of ELG candidates has been calculated separately for
seven consecutive 7\AA\ ($300$\,\kms) wavelength slices (Figure 
\ref{nompc}), under the gross assumption that all the candidates are real. 
An important point (see Section \ref{sec:ttfsetup}) is that the total 
wavelength range covered at any given pixel is different at different
locations in the field, and so certain parts of the field 
are not sampled at some wavelengths. Simply because of the geometry 
of the optical setup used, more of the field is sampled to the blue 
than to the red of the quasar redshift. Figure \ref{waveimage} 
shows the spatial limits of the areas sampled in each of the 
seven wavelength slices discussed above. Obviously, therefore,  
the total volume sampled per bin varies with wavelength. 

With allowances for volume sampling, Figure \ref{nompc} shows that 
the typical space density
of TTF ELG candidates found around MRC\,B0450--221 is of order
0.03--0.05\,Mpc$^{-3}$. There is a hint of an excess of objects
within $\pm900$\kms\ of the quasar,  but because of the small number 
of objects in each bin, this trend is not formally significant.

\subsection{\oii\ equivalent widths and star-formation rates}

Equivalent widths and line luminosities have been 
calculated from the TTF data as part of the analysis, assuming that 
enhanced emission in the narrow TTF bands is indeed due 
to \oii\ at $z=0.9$. Line luminosities above the
continuum magnitude (as estimated above) were calculated simply by adding
the excess flux in the three bins centred on the line peak.  This process 
may underestimate line flux if the peak does not occur in the 
central bin or if the lines are broad.
For objects where the continuum was not detected at all in the 
TTF data by FOCAS, 
an upper limit of $I(AB)=22.0$ 
(equivalent to the detection threshold) was assumed. 
The objects with the highest equivalent widths necessarily had
faint continua (at or below the detection limits) and so are subject to 
large uncertainties --- uncertainties of one or two magnitudes 
in the continuum level of objects with $I(AB)= 22$--23 will lead to 
uncertainties in equivalent widths of factors of 2--6. 

\begin{figure*}
%%\centerline{\psfig{file=figew.ps,bbllx=
\centerline{\psfig{file=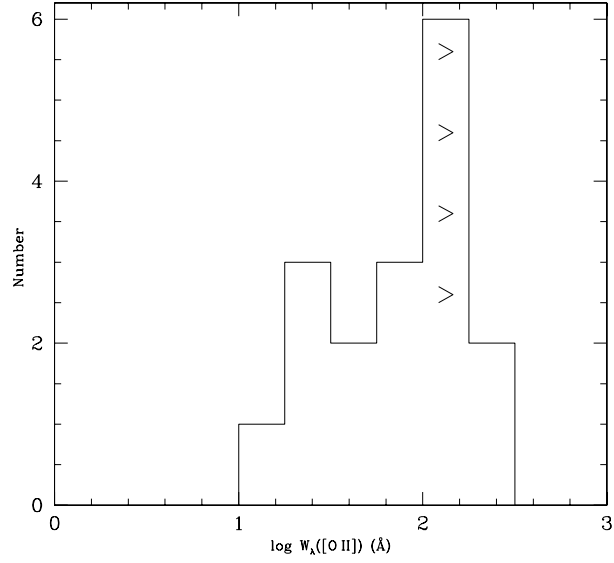,bbllx=
18pt,bblly=145pt,bburx=583pt,bbury=701pt,clip=,height=8.5cm} }
\caption[fig11]{\footnotesize
Histogram of measured \oii\ equivalent widths, $W_{\lambda}$, 
as estimated by the TTF analysis
for $>2.5\sigma$ ELG candidates with continua fainter than $I(AB)=21$. 
Lower limits  are marked for objects 
where the continuum level was undetermined
(assuming $I(AB)\approx 22$) --- the uncertainty in \ew\ 
will be large for these objects.
\label{ews}
}
\end{figure*}

For the 17 candidates ($>2.5\sigma$) 
fainter than $I(AB)=21$, equivalent widths \ew\ 
in the range 14--275\,\AA\ are measured (see Figure \ref{ews}), 
or 7--145\,\AA\ in the restframe if the galaxies lie at $z=0.9$.
Three objects have lower limits within the  
range \ew$=110-160$ \AA, and two more have limits \ew$>20$ and \ew$>50$\,\AA. 

\begin{figure*}
%%\centerline{\psfig{file=figsfr.ps,bbllx=
\centerline{\psfig{file=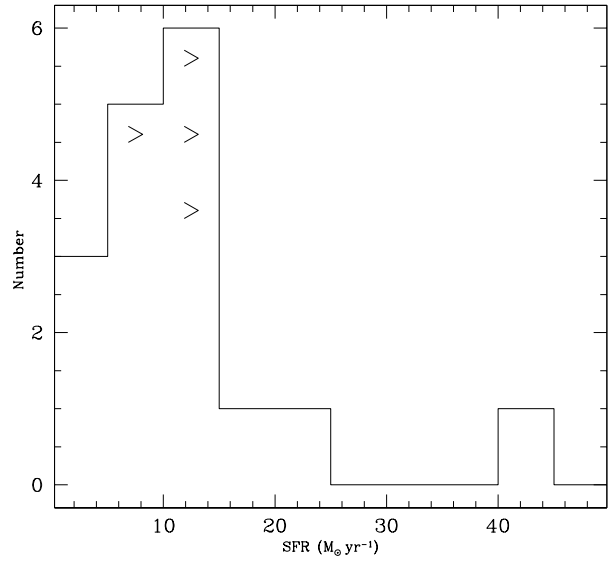,bbllx=
18pt,bblly=145pt,bburx=583pt,bbury=701pt,clip=,height=8.5cm} }
\caption[fig12]{\footnotesize
Histogram of star-formation rates inferred from \oii\
line luminosities taken from the TTF analysis
(for  $>2.5\sigma$ ELG candidates fainter than $I(AB)=21$mag)
using the conversion of Gallagher et al. (1989).
Lower limits  are marked for objects 
where the continuum level was undetermined
(assuming $I(AB)\approx 22$).
\label{sfr}
}
\end{figure*}

The corresponding line luminosities, assuming the objects lie at $z=0.9$,
all exceed $10^{41}$ erg\,s$^{-1}$. Using the 
conversion of Gallagher, Bushouse \& Hunter
(1989) from \oii\ line luminosity ($L_{[O II]}$ in erg\,s$^{-1}$)
to star-formation rate ({\it SFR} in \msunyr) ---
{\it SFR}\,$ \approx L_{[O II]} \times 10^{-41}$ ---
equivalent star formation rates of 1--40\,\msunyr\ are 
implied (see Figure \ref{sfr}). As well as uncertainties
arising from redshift assumptions and measurement, 
the empirical conversion from 
$L_{[O II]}$ to star-formation rate is itself uncertain by a factor
of a few due to intrinsic scatter in the samples used to calibrate
the relationship against other more direct star-formation 
indicators (such as UV luminosity and hydrogen recombination lines). 
We note that these indicators are all highly model dependent
and include gross assumptions about metallicity, IMF and 
their redshift evolution as well as uncertainties in basic properties of 
gas, dust, ionisation  and extinction in star-forming galaxies 
at $z\sim1$
(see Gallagher et al. 1989; Kennicutt 1992; Hammer et al. 1997
and the review by Kennicutt 1998).

\begin{figure*}
%%\centerline{\psfig{file=newbins.ps,bbllx=
\centerline{\psfig{file=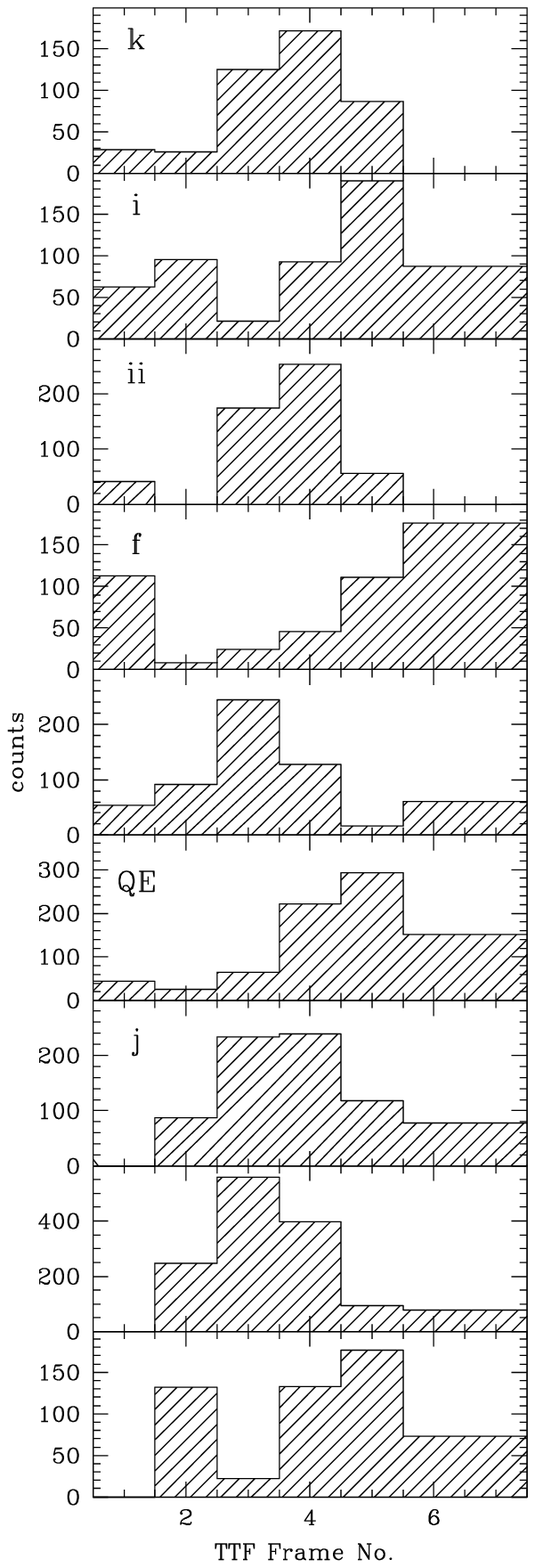,bbllx=
26pt,bblly=178pt,bburx=200pt,bbury=692pt,clip=,height=18cm} }
\caption[fig13]{\footnotesize
Histograms showing counts corresponding to the aperture
photometry measured on each TTF Frame (F1-F7) for the nine
ELG candidates with equivalent widths greater than 70\AA\
and fainter than $I(AB)=21$ in continuum. Objects which 
were not detected by FOCAS in a given image are assigned zero
counts in that bin.
The sky rms is 31 counts on this scale. 
The nine objects appear in order of Dec, from north to south. 
Ones which were targetted spectroscopically are
labelled (as in Figure \ref{mosids}), plus the 
quasar extended emission (QE).
\label{bins}
}
\end{figure*}

\subsection{High equivalent-width ELG candidates}
\label{sec:bestids}

We now consider only the strongest ELG candidates, nine objects
selected to have measured equivalent widths greater than 70\,\AA, 
and satisfying the other criteria  of $I(AB)>21$, and peak flux exceeding 
the $2.5\sigma$ level at wavelength longward of 7050\AA). In other words, 
this cutoff selects
objects having more than 60\% of their total counts recorded in
only one to three adjacent TTF frames, i.e. those faint objects 
most likely to possess the strongest emission lines. 
The counts measured in each TTF frame for these objects are
shown in Figure \ref{bins}. 
The emission-line strengths implied for these nine objects translate to 
star-formation rates of typically 10\msunyr\ and higher,  
peaking in the wavelength range 7061--7087\,\AA. 

Figure \ref{bestim} shows the location of the strongest TTF emission-line
candidates (circled) superimposed on the coadded TTF image.
Most lie in a group about 200--700\,kpc to the west of the quasar. 
Three additional objects with equivalent widths $50<$\ew$<70$\AA\
lie nearby, and are also marked in Figure \ref{bestim} (squares). 
The group of strong TTF candidates is also identified on the 
broad $I$-band image (note the smaller field of view) in 
Figure \ref{ttfiim}.
With the exception of the quasar extended line emission, 
all candidates in this region are clearly
identified with galaxies visible on the $I$-band image.

The tightest grouping of 5 ELG candidates contains objects peaking 
slightly blueward but close to the quasar redshift, in the 
7063--7073\AA\ range. Figure \ref{waveimage} clearly shows 
that these wavelengths are well sampled across the field, 
and so the clustering of objects is likely to be real rather than 
merely an artifact of the wavelength coverage. 
The wavelength gradient across the field will, however, restrict
the sampling of objects to the red of the quasar in the northern 
third of the field only, although only one such object was 
found in the southern part of the field. 

\begin{figure*}
%%%\centerline{\psfig{file=ttfids.ps,bbllx=
\centerline{\psfig{file=baker.fig14.ps,bbllx=
37pt,bblly=144pt,bburx=570pt,bbury=647pt,height=10cm} }
\caption[fig14]{\footnotesize
TTF ELG candidates with equivalent widths greater than 70\AA\
(circles) and 50$<$\ew$<70$\AA\ (squares) identified on the 
$7' \times 7'$ TTF field (co-added, smoothed TTF image shown). 
The quasar is identified with a `$+$'. The object circled
just to the NW of the quasar is assumed to be due to extended 
line emission.
}
\label{bestim}
\end{figure*} 

\begin{figure*}
%%\centerline{\fbox{\psfig{file=figiim.ps,bbllx=
\centerline{\psfig{file=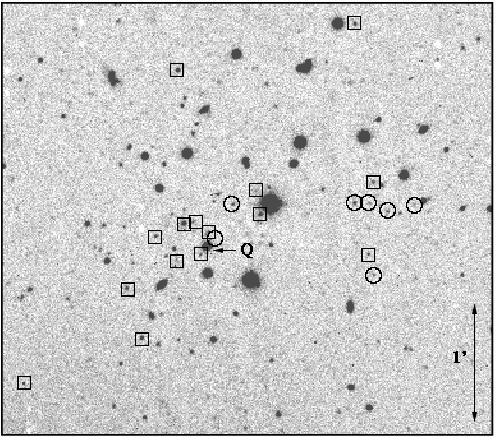,bbllx=
36pt,bblly=134pt,bburx=567pt,bbury=666pt,height=10cm} }
\caption[fig15]{\footnotesize
TTF ELG candidates (circles) and red galaxies (squares)
plotted on the $I$-band image. 
The TTF candidates have \ew$>70$\AA, and the 
red galaxies have $I>20$ and $V-I>2.2$ (see text).
The quasar is identified by an arrow. 
The quasar extended emission-line region is not detected
in the broad-band $I$ image. 
Note the enhancement of faint objects in the $I$-band image
within an arcminute or so of the quasar. 
The small white patches discernible in the background are artifacts 
from star residuals in the flat-field image used.
North is up, East left.
}
\label{ttfiim}
\end{figure*}

\subsection{Spectroscopy of selected objects}
\label{sec:mos}

In a weather-affected observing run, we managed to obtain
spectra for 12 galaxies within $2'$ of the quasar, 
including five TTF ELG candidates from the list above 
with estimated \ew$>70$\AA. 
Figure \ref{mosids} identifies the spectroscopic
targets (labelled $a$ to $l$; bright star $m$ lies to the 
south of the region shown; 
two objects were observed in slit $i$, and are denoted by 
$i$ and $ii$). The results are summarised in Table \ref{tabmos}.
Spectra of five objects, including those with emission lines
at $z=0.9$ and a lower redshift example, are shown in Figure \ref{mosspec}.

Of the five ELG candidates targetted, \oii\ at $z=0.9$ was detected 
clearly in three (objects $f, ii, j$ in Figure \ref{mosids}). 
The spectrum of object $k$ is consistent with weak \oii\ emission
at $z=0.9$, but noise makes the line identification uncertain.
The fifth candidate yielded a lower redshift identification
(object $i$ showed strong \oiii\ and \hb\ at $z=0.351$ 
outside the TTF window), 
and hence its detection by TTF was spurious.
In the three spectra with clear \oii\ detections, 
continuum emission is present blueward of 
4000\AA\ (rest), consistent with star-formation as the source 
of the \oii\ emission (and ruling out \lya\ emission at higher redshift,
where a continuum break due to absorption blueward of the line
would have been expected). 
The \oii\ equivalent widths estimated from the TTF magnitudes
agree well with those measured in two cases where the continuum magnitude
is well determined: for objects $f$ and $j$ TTF estimates 
\ew$\approx130$\AA\ and 160\AA\ and the spectra give
\ew$=140\pm50$\AA\ and $130\pm50$\AA\ respectively (in the observed frame). 
For the fainter object,
$ii$, the spectroscopic equivalent width, \ew$=70\pm25$\AA, is
smaller than the estimation based on TTF magnitudes, \ew$>160$\AA\ 
(which assumed the continuum lies at the magnitude 
limit $I(AB)=22$). However, this is still within 
the expected photometric uncertainties in the TTF data
(factors of 2--6 at $I(AB)=22$--23). As object $ii$ was 
essentially observed serendipitously, falling right on the edge of
mini-slit $i$, it is almost certainly affected by slit losses.
These losses could affect the line-to-continuum ratio in an 
extended object if the star-formation is patchy 
and/or highly nucleated. For all three identified 
\oii-emitters the spectroscopic and TTF peak wavelengths agree 
to within 3--13\AA\  (i.e. one to two TTF bins), as expected.

Of the remaining multi-object spectroscopic targets 
(chosen more-or-less at random), 
galaxies $a, b$ and $e$ showed strong emission lines at lower redshifts
(outside the TTF window),
and the remaining objects ($c,d,g,l$) lacked  
identifiable emission or absorption features in their noisy spectra.
Object $d$ is one of the reddest objects in the field
($V-I = 2.95$) and so is a candidate for a passively-evolving 
elliptical galaxy at $z=0.9$. However, its spectrum was too 
noisy to confirm a redshift and further observations are being sought. 

For the quasar itself, the equivalent width of the \oii\ emission 
estimated by TTF was 19\AA\ (observed frame, in 3\arcpt 6 diameter circular 
aperture). This compares with \ew=$10\pm 2$\AA\ measured in a 
recent longslit spectroscopic observation using a 1\arcpt 5 wide slit
(de Silva et al., in preparation). 
Because of the presence of strong extended line emission
close to the quasar nucleus (\ew$>200$\AA), 
it is not surprising that the equivalent width 
measured in the larger TTF aperture is somewhat higher. 	

\begin{figure*}
%%\centerline{\psfig{file=mosids.ps,bbllx=
\centerline{\psfig{file=baker.fig16.ps,bbllx=
108pt,bblly=197pt,bburx=505pt,bbury=590pt,height=8cm} }
\caption[fig16]{\footnotesize
Identification of spectroscopy targets 
(circled) on the co-added TTF image 
(cropped to $135'' \times 135''$, East left, North up).
The targets are labelled $a$ to $l$ (fiducial star $m$ not shown). 
Two galaxies were observed in slit $i$, and have been
denoted $i$ and $ii$. The quasar is marked `Q'. 
The dispersion direction ran north-south.
\label{mosids}
}
\end{figure*}

\begin{figure*}
%%\centerline{\psfig{file=mosspec.ps,bbllx=
\centerline{\psfig{file=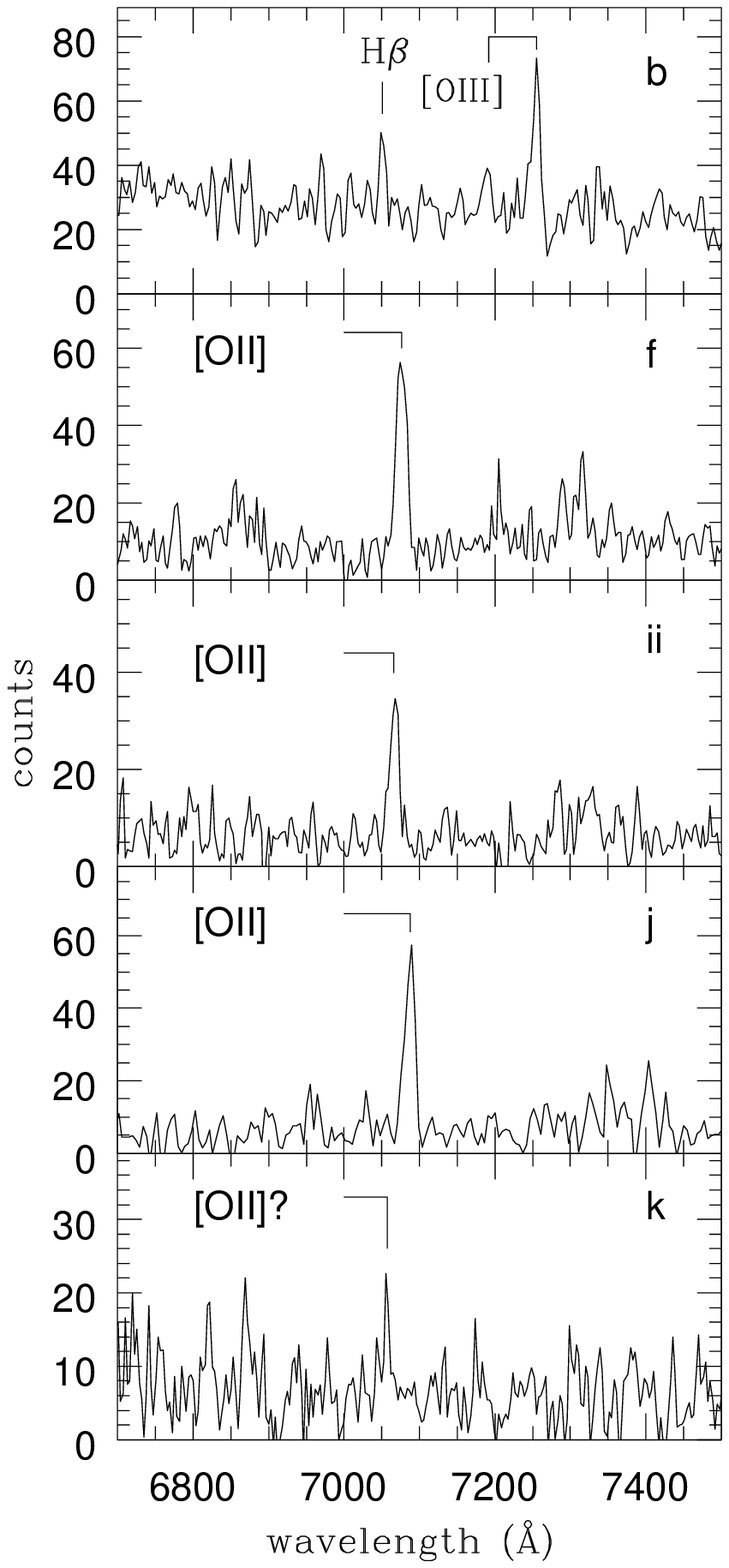,bbllx=
41pt,bblly=163pt,bburx=289pt,bbury=692pt,clip=,height=15cm} }
\caption[fig17]{\footnotesize
Spectra of five galaxies  
($b,f,ii,j,k$ as identified in Figure \ref{mosids}). 
Object $b$ is at $z=0.449$ (note \hb\ and the 
\oiii$\lambda4959,5007$ doublet marked longward).
The other objects show \oii$\lambda3727$ at $z=0.9$ ($k$ is uncertain). 
\label{mosspec}
}
\end{figure*}

\section{Clustering of red galaxies}
\label{sec:optir}

\subsection{Optical}

\begin{figure*}
\centerline{\psfig{file=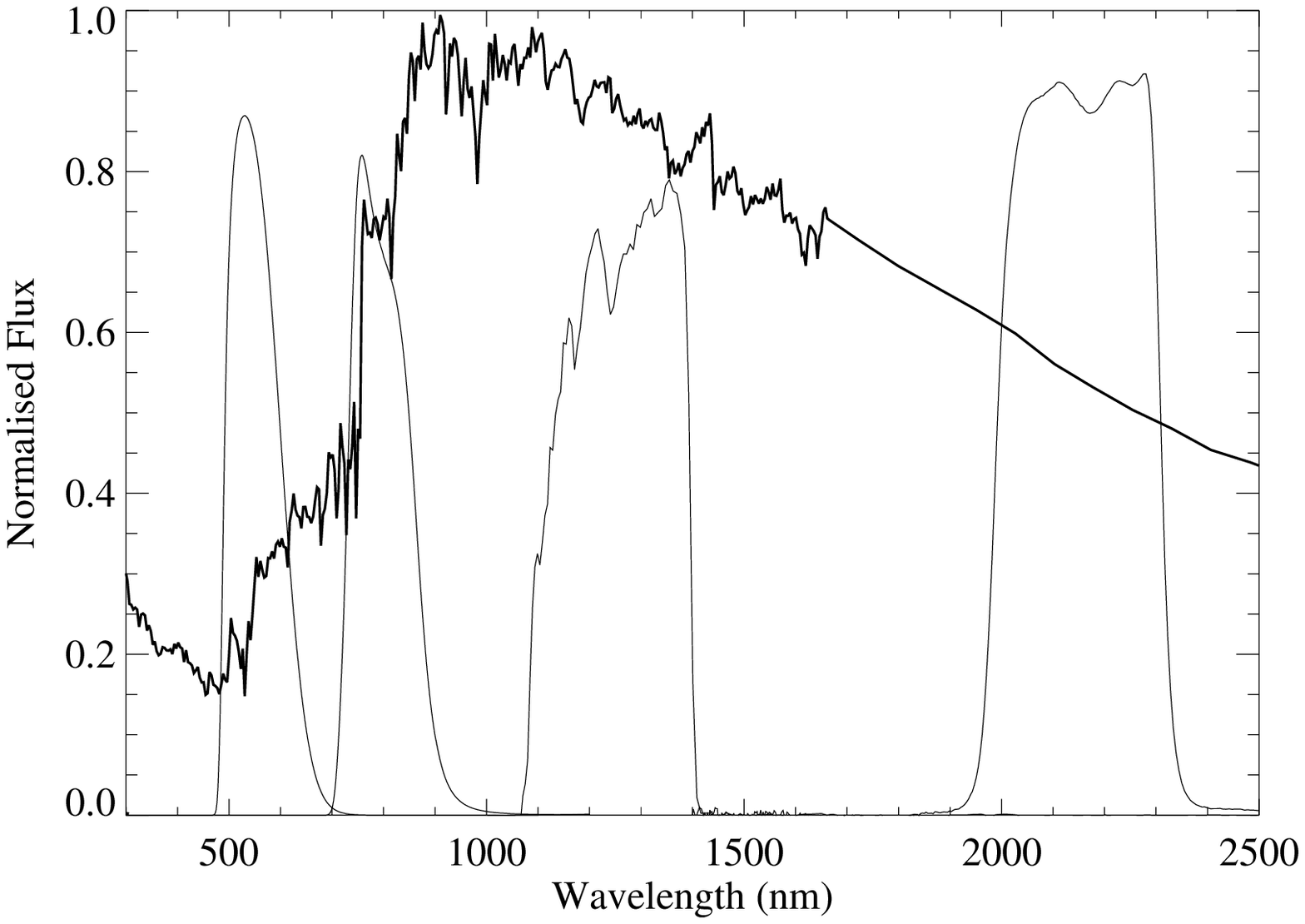,bbllx=
77pt,bblly=369pt,bburx=555pt,bbury=708pt,clip=,height=5cm} }
\caption[fig18]{\footnotesize
Typical spectrum ($f_{\lambda}$) of an elliptical galaxy, 
redshifted to $z=0.9$, 
that has evolved passively for 3Gyr
since its last burst of star formation (e.g. Fioc \& Rocca Volmerange 1997). 
Overlaid are transmission curves for the $V$, 
Gunn-$i$, $J$ and $K'$ filters. 
\label{sed}
}
\end{figure*}

\begin{figure*}
\centerline{\psfig{file=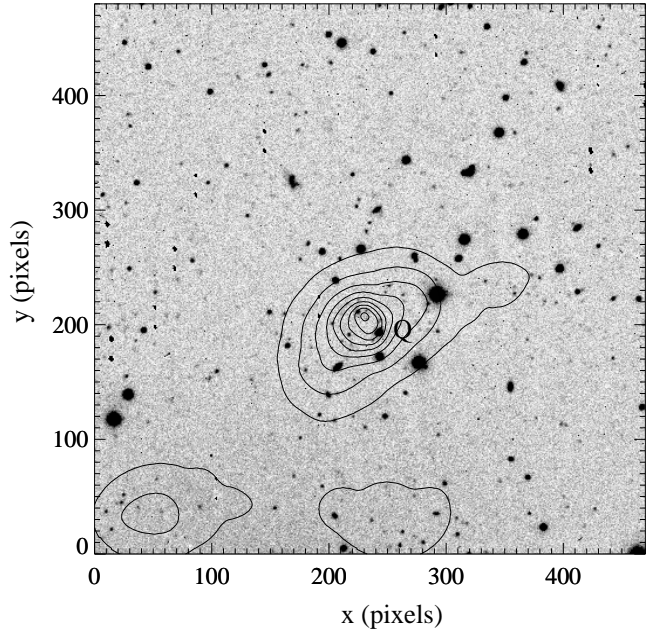,bbllx=
50pt,bblly=205pt,bburx=442pt,bbury=590pt,height=8.5cm} }
\caption[fig19]{\footnotesize
Contour plot showing clustering of red galaxies ($I>20$, $V-I>2.2$; 
see Figure \ref{ttfiim}) overlayed on the $I$-band image. 
The density of red objects (contours) has been 
quantified using a non-linear measurement to emphasise the clustering
--- the inverse of the average distance to the nearest five objects,
calculated for each red galaxy in the field. The contour levels 
represent the smoothed distribution of this quantity in increasing
steps of $1\sigma$, starting with the $3\sigma$ contour 
and peaking at $11\sigma$.
Clearly the greatest clustering of red objects occurs close 
to the quasar (Q), and also extends out towards the westward group
of ELGs. The scale is shown in pixels (0\arcpt 61). 
\label{redcont}
}
\end{figure*}

\begin{figure*} 
%%%%\centerline{\psfig{file=radred.ps,bbllx=
\centerline{\psfig{file=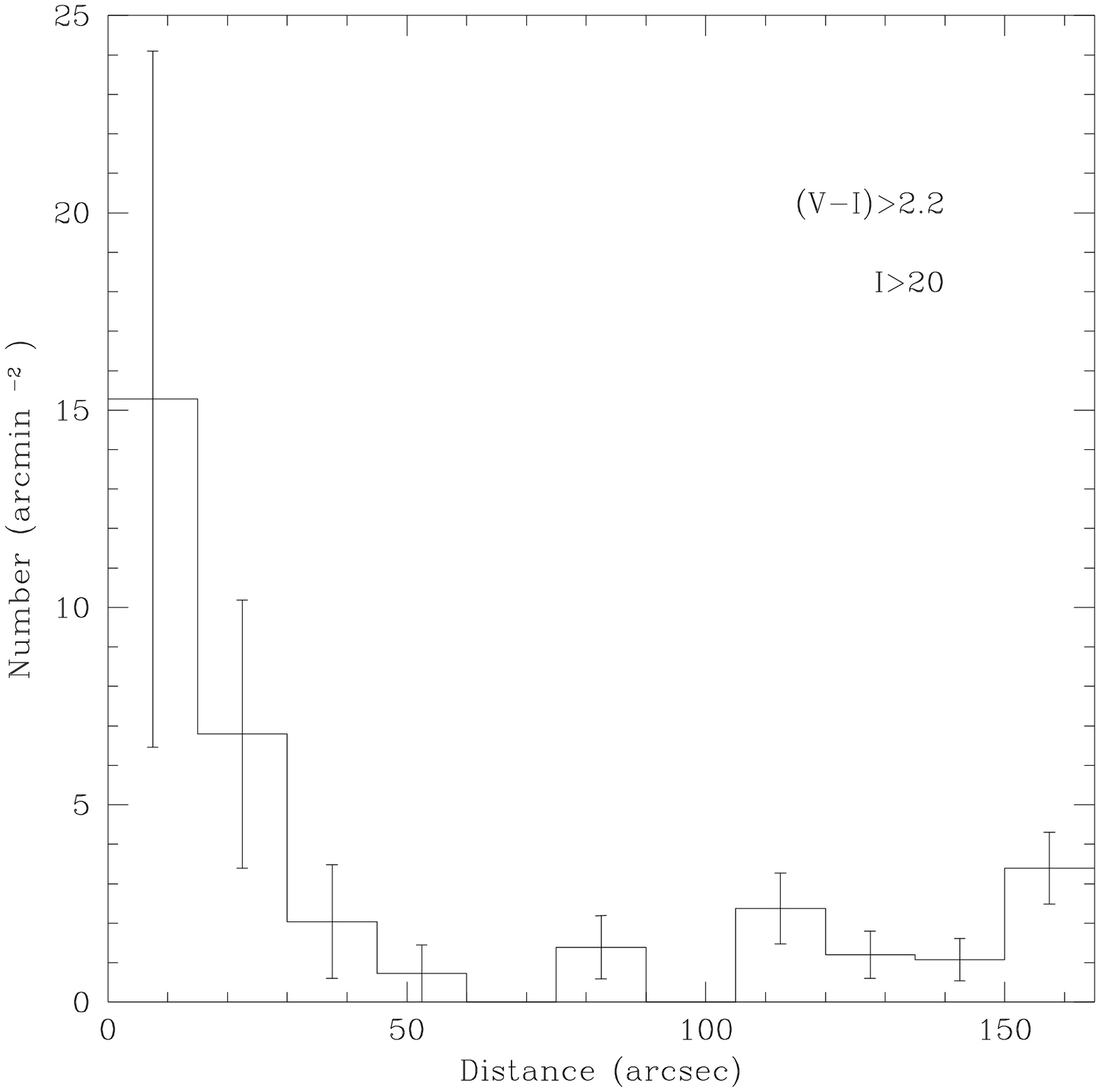,bbllx=
20pt,bblly=150pt,bburx=570pt,bbury=700pt,clip=,height=8.5cm}}
\vspace{1mm} 
%%%\centerline{\psfig{file=radblue.ps,bbllx=
\centerline{\psfig{file=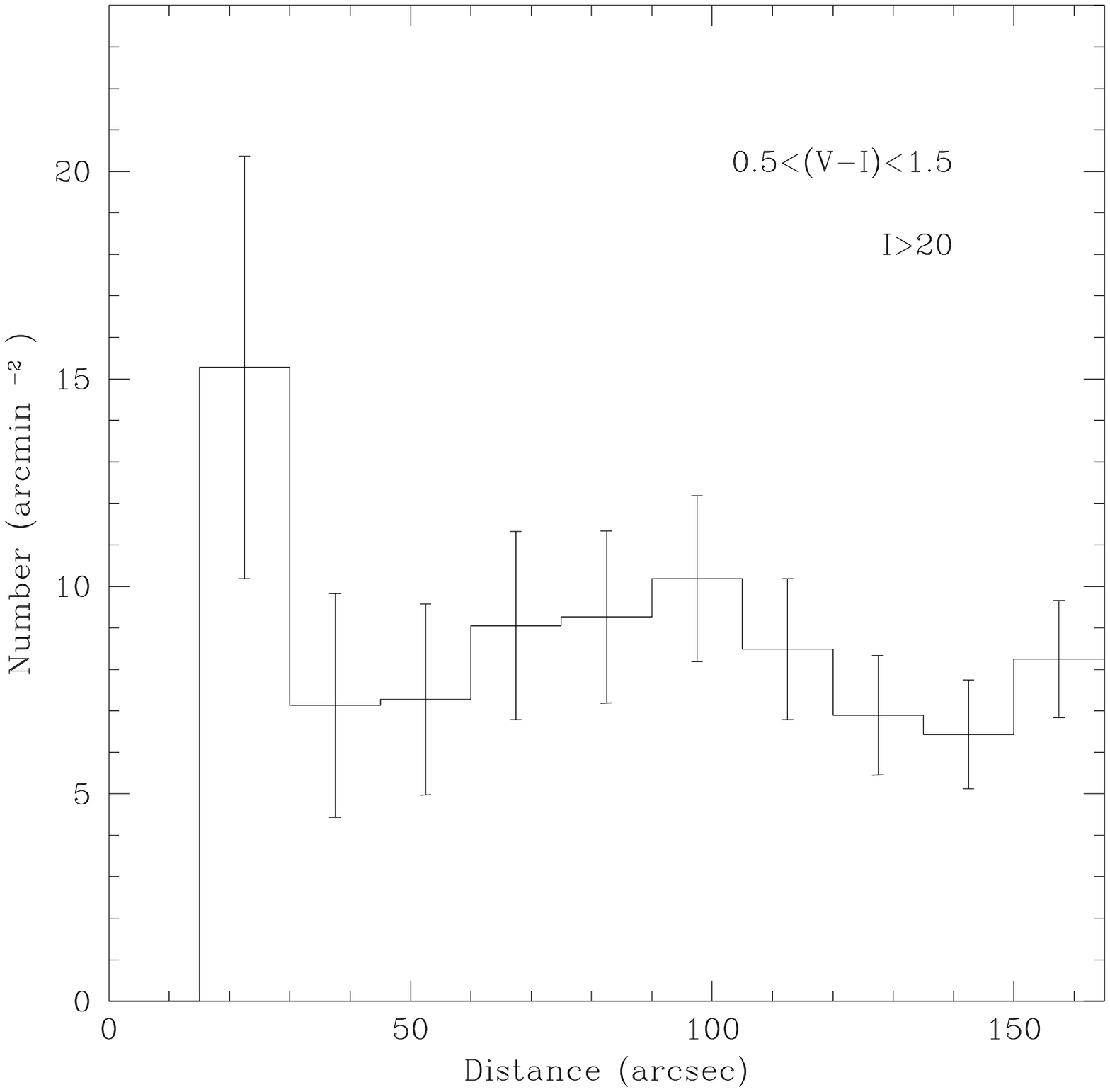,bbllx=
20pt,bblly=150pt,bburx=570pt,bbury=700pt,clip=,height=8.5cm}}
\caption[fig20ab]{\footnotesize 
{\it Top:} Surface density of faint objects ($I>20$) with red colours
($V-I>2.2$) as a function of distance from the quasar. 
Note the excess near the quasar.
{\it Bottom:} Surface density of faint objects ($I>20$)
with more typical colours ($0.5<(V-I)<1.5$; see Figure \ref{ivhist}) 
as a function of distance from the quasar. 
\label{radplots}
}
\end{figure*}

Because the $I$ and $V$ filters were chosen to 
straddle the redshifted 4000\,\AA\ spectral break at $z=0.9$, 
we expect objects with the reddest $V-I$ colours 
to include passively-evolving galaxies at $z=0.9$. 
To illustrate this, Figure \ref{sed} shows the filter 
transmission curves with a typical spectrum of an elliptical 
galaxy which has evolved passively for about 3 Gyrs overplotted.  
An overdensity of these red galaxies might signify
a cluster, as passive evolution is characteristic of ellipticals in 
dense cluster cores. Indeed, Figures \ref{ttfiim} and \ref{redcont}
show an overdensity of galaxies 
with $V-I >2.2$  in the field, centred near the quasar. 
Figure \ref{radplots} shows the  surface density of faint, 
red galaxies ($I>20$ and $V-I >2.2$) as a function of distance 
from the quasar. The red galaxies peak strongly close to the quasar,
enhanced relative to the surrounding density by a factor of order five
within an arcminute radius of the quasar.

\begin{figure*}
%%%%\centerline{\psfig{file=newivhist.ps,bbllx=
\centerline{\psfig{file=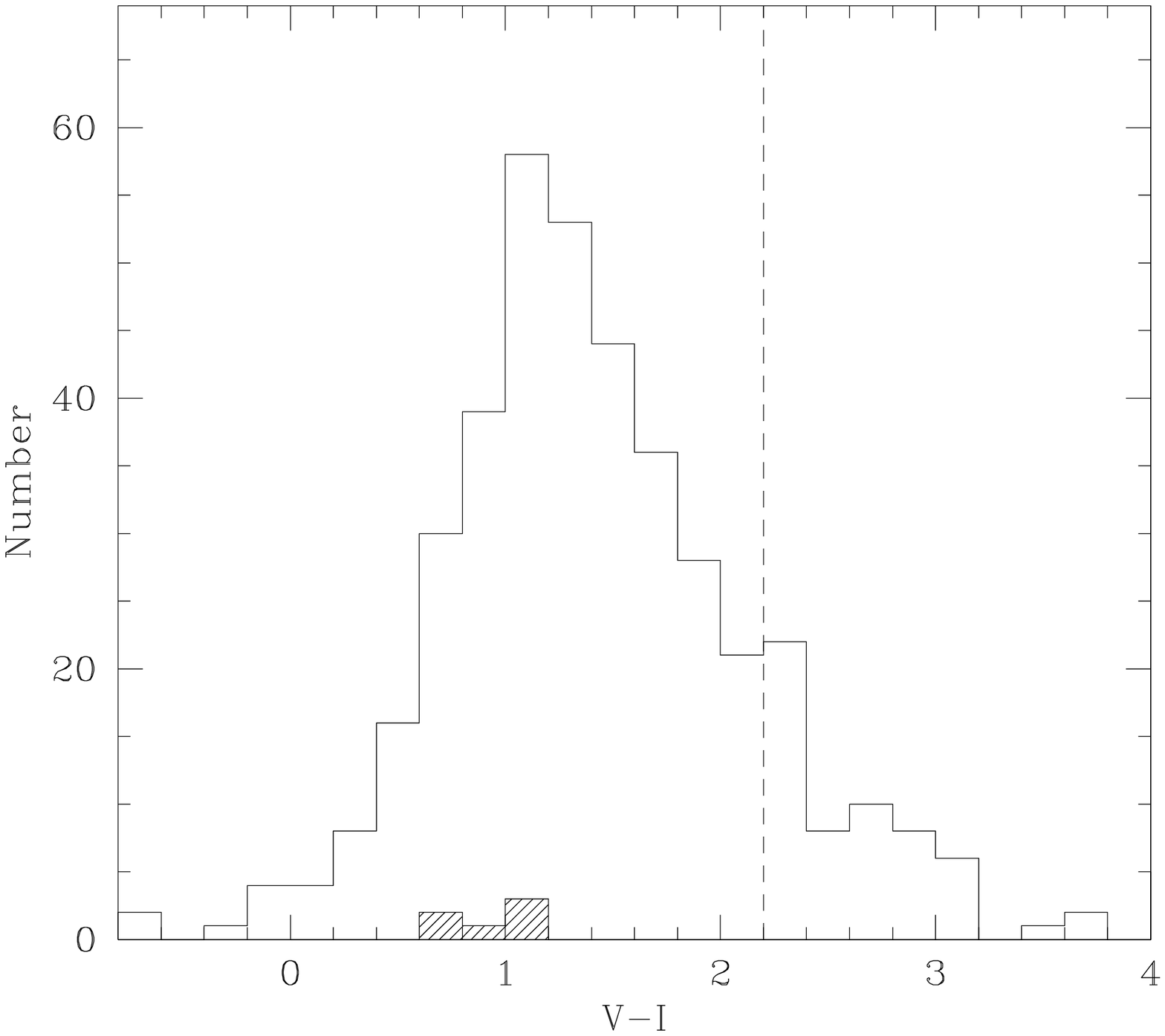,bbllx=
26pt,bblly=155pt,bburx=570pt,bbury=700pt,clip=,height=8.5cm}}
\caption[fig21]{\footnotesize 
Histogram of $V-I$ colours for all objects detected in
both $I$ and $V$ broad-band images.
TTF ELG candidates ($I(AB)>21$ and \ew$>70$\AA) are
also shown (shaded) for comparison, excluding three
without broad-band colours (two outside the field, plus the 
quasar extended line emission).
The dashed line marks the defining limit of the
red galaxies, $V-I \geq 2.2$.
\label{ivhist}
}
\end{figure*}

Also shown in Figure \ref{radplots} is the surface 
density of faint galaxies with average $V-I$ colours 
(namely $0.5<(V-I)<1.5$, and $I> 20$)  as a function of radial distance 
from the quasar. 
The distribution of $V-I$ colours for all galaxies in the field 
of MRC\,B0450--221 is shown in Figure \ref{ivhist}. The broad range 
of colours is comparable with that found in other field surveys 
(e.g. Hammer et al. 1997).
In contrast to the distribution for 
red galaxies, the surface density of blue galaxies remains 
approximately constant across the field, 
apart from a slight excess $25''$ from the quasar and a deficit of 
blue galaxies within $10''$ of the quasar itself. 
Within this typical colour range, the strongest TTF candidates 
(\ew$>70$\AA, shaded in Figure \ref{ivhist}) 
span a fairly small range, $V-I\sim 0.7$--1.1, 
about two magnitudes bluer than the red galaxy excess described above. 

The colours of the red galaxies are consistent with 
those predicted by spectral synthesis models (such as the PEGASE models 
of Fioc \& Rocca-Volmerange 1997) for elliptical galaxies at $z=0.9$, 
namely, those which have been evolving passively (SFR$<1$\,\msunyr) 
for $3$\,Gyr or more, following a 1-Gyr duration burst of star 
formation, and  including a modest amount of reddening 
($E_{\scriptstyle B-V} = 0.25$).
We note that the $V$ magnitude is very sensitive to any 
ongoing star formation, and so some spread of $V-I$ would be 
expected even for coeval galaxies. Spectroscopy with 8-m class telescopes
will ultimately be needed to 
probe the star-formation histories of individual galaxies in more detail. 

\subsection{Infrared}

\begin{figure*}
%%%%%\centerline{\psfig{file=radir.ps,bbllx=
\centerline{\psfig{file=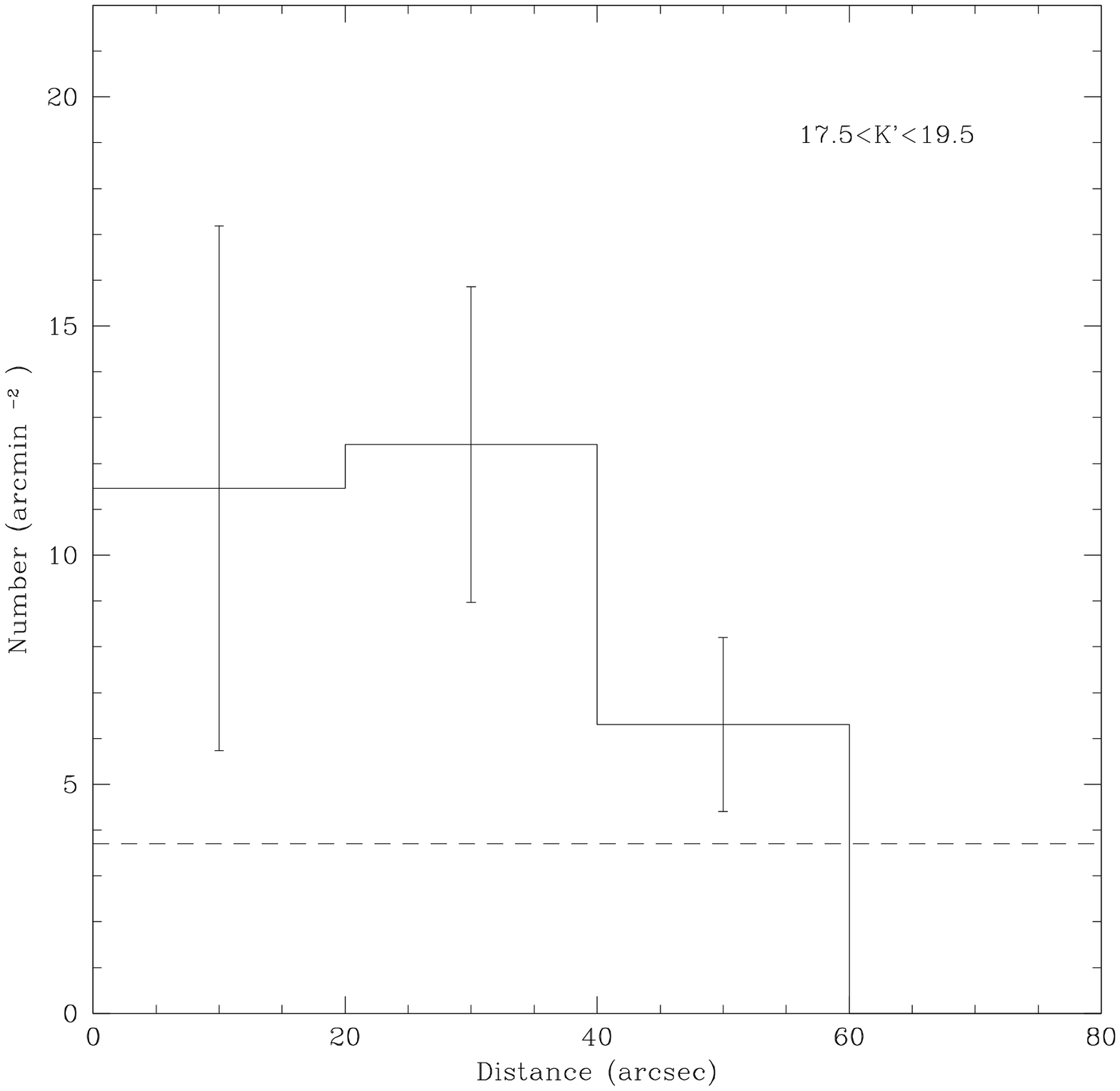,bbllx=
28pt,bblly=155pt,bburx=570pt,bbury=700pt,clip=,height=8.5cm}}
\caption[fig22]{\footnotesize 
Surface density of objects (arcmin$^{-2}$) detected in both
the $J$ and $K'$ images, and with $17.5<K' < 19.5$, plotted as a 
function of radius (arcsec) from the quasar. One sigma error 
bars are shown based on number of objects per bin. The dotted 
line indicates the expected level for this mag range 
taken from infrared field surveys 
in the literature (McCracken et al. 2000), with 10\%
uncertainty. Edge effects dominate beyond $80''$. 
\label{radir}
}
\end{figure*}

The near infrared images probe the region of the red galaxy excess 
near the quasar, and many of the reddest objects are 
detected in the infrared (see Figure \ref{four}). 
Figure \ref{radir} shows a plot of radial surface density
of the 28 infrared objects with $17.5<K'<19.5$ as a function 
of distance from the quasar. There is an obvious excess of 
objects within about $40''$ of the quasar 
(17/28 objects, or 12 arcmin$^{-2}$), 
which is consistent with a moderate richness cluster even 
though small number statistics limit its formal significance. 
For the same magnitude range, the $K$-band counts of McCracken 
et al. (2000) imply that the expected number density of objects in a 
random field  is only about 4 arcmin$^{-2}$. This number assumes
50\% completeness in the range $19.0<K'<19.5$, which is 
where the counts per unit magnitude in the MRC\,B0450 field
start to turn over (if no correction was made the field density 
would increase to only 5 arcmin$^{-2}$). 
The scale, $40''$, of this excess of infrared galaxies in our single field
is the same as that found on average by Hall \& Green (1998) in their
combined data for many quasar fields at $z=1$--2.

\begin{figure*}
%%%%%\centerline{\psfig{file=ikhist.ps,bbllx=
\centerline{\psfig{file=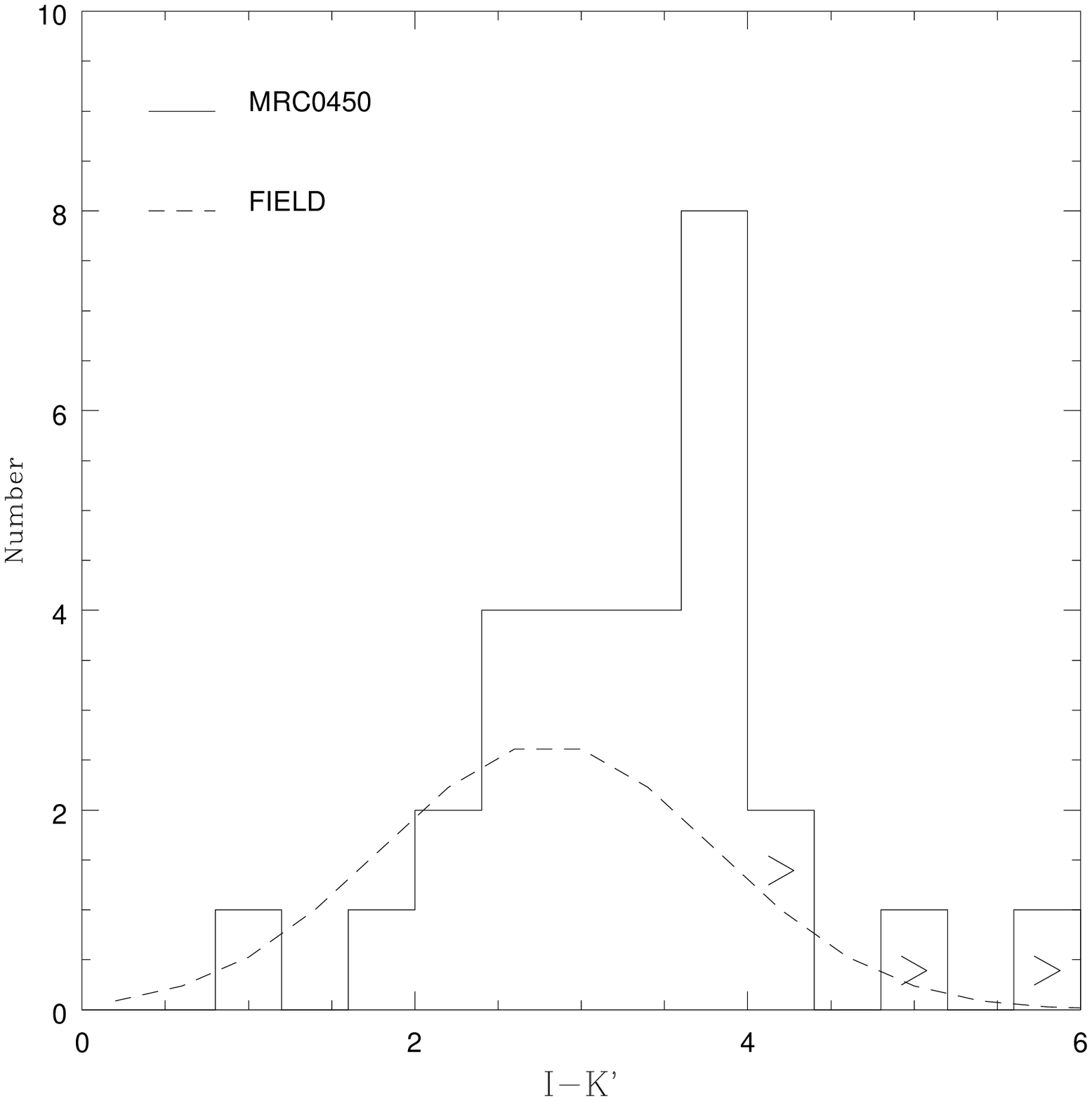,bbllx=
28pt,bblly=155pt,bburx=570pt,bbury=700pt,clip=,height=8.5cm}}
\caption[fig23]{\footnotesize 
Histogram of $I-K'$ colours for objects
detected in the infrared images with $17.5<K' < 19.5$.
The total area covers 4.5 arcmin$^2$ around the quasar.
Lower limits (objects below the detection theshhold in $I$)
are indicated by arrows. 
The dashed line shows the expected colour distribution 
based on infrared field surveys (Thompson et al. 1999), normalised to
the counts of McCracken et al. 2000 (see text).
Note the spike of objects with   $I-K'\approx 3.8$. 
\label{ikhist}
}
\end{figure*}

Figure \ref{ikhist} shows the distribution of $I-K'$ colours 
for the galaxies in the MRC\,B0450--221 infrared frame with $17.5<K'<19.5$.
As excess of objects is obvious, notably a spike of objects with 
$I-K'=3.8\pm0.2$. As shown in Figure 6 of McCracken et al. (2000),
the $I-K'$ colours of non-evolving galaxies are fairly insensitive to 
any ongoing starformation and increase smoothly  
with redshift out to $z\sim 2$. At $z\approx0.9$ they would 
indeed be expected to have colours of $I-K'\approx 3.8$.
For models including some recent star-formation, 
colours as blue as $I-K' \approx 2.5$
would be expected for galaxies at  $z\approx 0.9$, and this overall
range of colours agrees well with the colours of the galaxy excess in 
Figure \ref{ikhist}.
The equivalent distribution expected for randomly-selected
field galaxies has been approximated in Figure \ref{ikhist} by 
a Gaussian (Thompson et al. 1999), with mean $I-K'\approx 2.8$ 
(assuming $R-I\approx 1$ on average, Liu et al. 2000)
and FWHM 2.4 mags.  The Gaussian distribution was scaled 
using the $K$-band counts of McCracken et al. (2000), whose
$I-K'$ distribution, incidentally, is also consistent with 
this parameterisation. Looking now at the density of objects 
with the most extreme red colours, we see 4 objects with  
$I-K' \geq 4$ in Figure \ref{ikhist}, three of which have 
$18<K'<19$. For comparison, field surveys (McCracken et al. 2000)
find about $0.2\pm0.1$ objects arcmin$^{-2}$ with $18<K<19$ and $I-K \geq 4$,
whereas we see $0.7 \pm 0.4$ arcmin$^{-2}$ around MRC\,B0450--221. 

\begin{figure*}
%%%%%\centerline{\psfig{file=jkhist.ps,bbllx=
\centerline{\psfig{file=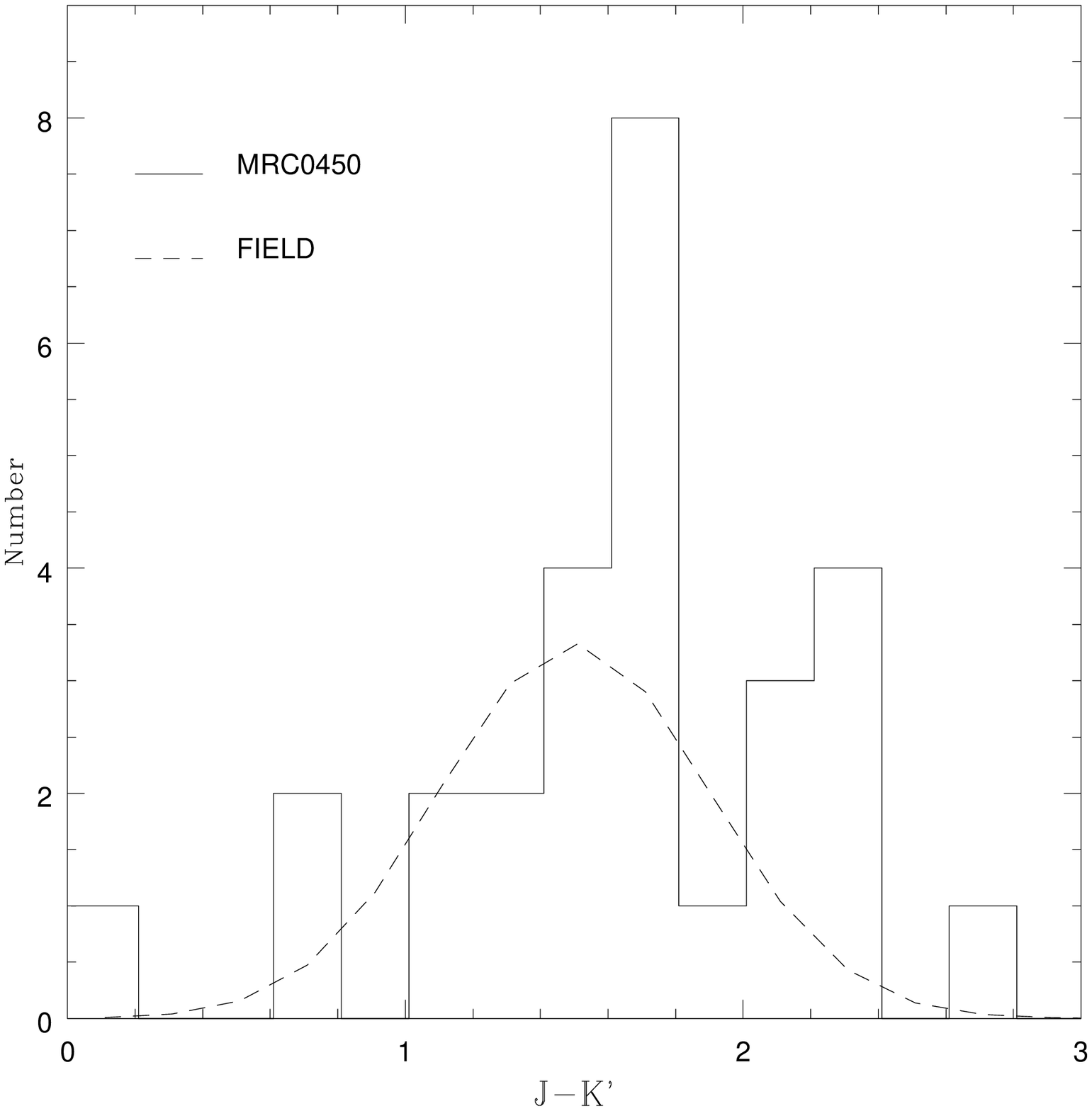,bbllx=
28pt,bblly=155pt,bburx=570pt,bbury=700pt,clip=,height=8.5cm}}
\caption[fig24]{\footnotesize 
Histogram of $J-K'$ colours for objects
detected in the infrared images with $17.5<K' < 19.5$.
The dashed line shows the expected distribution 
based on infrared field surveys (Saracco et al. 1999; see text).
Note the spike of galaxies with $J-K'= 1.7 \pm0.1$, and excess
with $J-K'=2.0$--2.4, relative to the field.
\label{jkhist}
}
\end{figure*}

Similarly, a histogram of $J-K'$ colours is presented in Figure \ref{jkhist}. 
The Gaussian comparison distribution of colours for an infrared-selected 
field sample with $K=17.5$--19.5 was taken from Saracco et al. (1999), 
and has mean $J-K=1.5$ and FWHM 0.9~mags. Again, an excess of
galaxies is indicated, including very red ones 
with $J-K'=2.0$--2.4 and a spike at $J-K'=1.7\pm0.1$. This spike is
only seen for objects with $18 < K' \leq 19$, where 9/16 objects have 
$J-K'=1.7 \pm 0.2$, and is not obvious for objects with $K'>19$.
These colours are again typical of
passively-evolving ellipticals at $z=0.9$ 
(Pozzetti \& Mannucci 2000).

\section{Discussion}
\label{sec:disc}

The optical and infrared imaging data themselves provide strong evidence
that the quasar MRC\,B0450--221 lies in a cluster at $z=0.9$, as
marked by an excess  of red galaxies within an arcminute of the quasar
with colours ($V-I>2.2$, $I-K' \approx 3.8$, $J-K'\approx 2$) 
expected for cluster 
ellipticals at $z=0.9$. At this redshift, the red galaxy overdensity
has dimensions of a few hundred kpc in diameter, which is typical of 
a cluster core. Similar arguments for excesses of red galaxies have 
been used to identify a number of clusters now at high redshift, 
including some around AGN (Stanford et al. 1997; Hall \& Green 1998).

The TTF observations provide another insight into the star-formation 
properties of this high-redshift system. 
With TTF, we effectively surveyed a volume of 420 $h_{50}^{-3}$Mpc$^{3}$
around MRC\,B0450--221 for emission-line objects, 
finding initially nine candidates  
with inferred star-formation rates exceeding about 10\,\msunyr. 
We note that despite their high star-formation rates, these 
detections include objects with extremely faint continuum levels, 
$V=24-25$ and $I>22$, which would have been missed by all but the 
deepest broad-band and spectroscopic surveys. 
At least three have been confirmed unequivocally by follow-up spectroscopy 
to be line-emitting galaxies at $z=0.9$.

To establish whether the number of detections is as expected, 
we must address the completeness of the TTF survey. To do this we
consider (i) statistical incompleteness in the detection of 
faint objects (4/9 candidates have average magnitudes below 
the completeness limit of $I(AB)=21.5$,  suggesting that others were missed), 
(ii) contamination from low-redshift interlopers and 
(iii) spurious  detections. The following arguments illustrate the 
relative importance of these factors, although ultimately they are
subject to large uncertainties being based on data for a single field.

In Figure \ref{complete}, some 36 objects lie in
the range $I(AB)=21.0$--21.5 (the bin just brighter than the 
completeness limit for the co-added TTF dataset),
whereas only nine lie in the next bin $I(AB)=21.5$--22.0. 
The slope of the cumulative 
counts per half-magnitude bin is approximately 1.25 up to the 
completeness limit,  $I(AB)=21.5$, so it follows that of order
44 objects would have been expected in the magnitude range 
$I(AB)=21.5$--22.0. Therefore, we estimate that 
as many as 80\% of objects may remain unidentified in this range. 
Of the ELG candidates themselves, 4/9 were below the completeness limit,
suggesting that (if the same ratio holds) 
$\sim 16\pm6$ similar line emitters might have been missed by 
the detection algorithm. Therefore, including the brighter ones, our
total number of nine ELG candidates may be incomplete by 50--70\%
due to missing faint candidates. Deeper observations should pick 
these up in future.

In the wavelength window targetted by TTF, emission lines from 
lower-redshift galaxies, such as \oiii\ and \ha, 
are also expected to be picked up.
The strongest expected line is \ha\ which would be 
detectable at $z\approx 0.1$; the TTF survey samples a volume of
only 37 $h_{50}^{-3}$Mpc$^{3}$ at $z \sim 0.1$. At $z\approx 0.1$,
a line with the same flux as a 10\,\msunyr\ \oii-emitter at $z=0.9$
would have a luminosity of about 100 times less, and the \ha\ line is
also typically 2--5 times stronger than \oii\ in starforming galaxies 
(Kennicutt 1992). Therefore, we would need to integrate contributions from 
\ha-emitters  down to $\sim0.05$\,\msunyr\ at $z\approx 0.1$ 
to compare with the strongest \oii-emitters ($>10$\msunyr) at $z=0.9$.
As the typical space density of $z=0.1-0.4$ galaxies with 
0.1--5 \msunyr\ is approximately constant per logarithmic
interval of 0.3 in star-formation rate, at 
about $3\times10^{-3}h_{50}^{3}$\,Mpc$^{-3}$ 
(Cowie et al. 1997) and falls off sharply above 5\,\msunyr, 
the contribution from \ha-emitters to our data is expected 
to be small, a rate of only $\sim 0.5$--1
object on average in the volume sampled. The number of expected 
\oiii\ emitters is harder to estimate because \oiii\ strength is only 
poorly correlated with star-formation rate.
However, assuming \oiii\gl5007 is similar to \oii\ in strength, 
$\sim 1$--2 objects are expected in  the 240 Mpc$^{3}$ volume sampled 
at $z\approx 0.4$. Similarly, another 0.5--1 \oiii\gl4959-emitters and
0.5--1 \hb-emitters might be detected at $z\approx 0.4$. 
Contamination by low-redshift 
objects on the basis of lines much weaker than \oii\ is negligible,
as only the strongest star-forming galaxies ($>10$\msunyr)
could contribute and their space densities  
are insignificant  at $z<0.8$ 
(according to the data of Cowie et al. 1997). 
Therefore, in total, only about 3--5 low-redshift field 
interlopers would be expected in a field of this area,
although these estimates are very sensitive to the assumed 
line strengths relative to \oii\ and large uncertainties 
(at least a factor of two) in the Cowie et al. data. 
The number of expected interlopers in the TTF dataset, however, 
will be lower than this because of the restricted magnitude range of
our detections, $22.5<I(AB)<21$, and incompleteness. 
So, reducing the number by a conservative factor of two, 
we realistically expect only of order one or two objects
(or less) to be low-redshift detections based on lines other than \oii.

Regarding spurious detections, 
the multi-slit spectroscopy shows that at least 
one candidate is spurious (maybe two if we include object $k$)
and therefore, by extrapolation, we can estimate that roughly
20--40\% of the TTF detections may also be spurious.

Cowie et al. (1997) find an average space density of
galaxies with star-formation rates
greater than 10\,\msunyr\ of $6.2 \times 10^{-4}$ $h_{50}^{3}$Mpc$^{-3}$
at $z\approx0.8-1.6$ (based on 27 galaxies, so an uncertainty of about 20\%). 
This translates to an expected number of 0.3 
line-emitting field galaxies in the TTF field, which is again 
probably an overestimate at $z=0.9$ as there are essentially no objects with 
lines this strong in the $z<0.8$ bins. Already, the 
three spectroscopically confirmed \oii-emitters themselves
suggest that we are seeing ELGs in numbers greater than expected in
the average field, and the significance of this excess will
only increase as other candidates are confirmed. We note that the
methods of selection ($B<24.5$) and estimation of SFR (luminosity 
at 2500\AA) used in the Cowie et al. redshift survey differ from our 
study. Although, Cowie et al. claim their continuum-based SFRs
are consistent with measured \oii\ strengths.

As shown in Figure \ref{bestim}, the ELGs with 
the largest equivalent widths (including spectroscopically-confirmed 
galaxies $ii,j$ and possible detection $k$) 
cluster spatially, most lying in a
group which is offset to the west of the quasar by 200--700\,kpc. 
This distance is similar to the characteristic radius in 
low-redshift clusters where the blue star-forming galaxy population 
begins to dominate the outer regions (Balogh \etal\ 1997; Morris et al. 1998).
The line strengths and equivalent widths 
(restframe equivalent widths from a few \AA\ to 140\AA\ 
and inferred star formation rates of 1--40\,\msunyr) of the ELGs
are consistent with other observations of field galaxies at 
redshift $z\sim 0.9$
(Cowie et al. 1997; Hammer et al. 1997).

Altogether, the environment of MRC\,B0450--221
is consistent with a moderately-rich cluster at $z\approx 0.9$, plus
a peripheral group of gas-rich galaxies being 
accreted from the surrounding overdense field.
The characteristics of the system are reminiscent of
other examples of merging clusters, including 
the red galaxy distribution being somewhat elongated 
(similar to the $z=0.8$ ROSAT-detected cluster RX\,J1716.6+6708; 
Henry \etal\ 1997; Clowe et al. 1998), and the presence
of an offset group of strongly star-forming galaxies 
(e.g. also a feature of the low-redshift cluster 
containing the powerful radio source Cyg~A; Owen et al. 1997). 
It is interesting to note that the system as a whole appears to
be elongated in roughly the same direction as the quasar radio axis;
observations of more fields will allow us to test whether this is 
coincidental or a feature of quasar environments.
With few spectroscopic redshifts available, 
the velocity dispersion of the  
MRC\,B0450--221 system is not constrained by the current data;
velocity dispersions of $\sim 800$\,\kms\ are seen 
typically in comparable systems such as low-redshift Abell Class O-I clusters 
(Dressler \& Shectman 1988; Zabludoff \etal\ 1993; Bower \etal\ 1997)
and clusters around some other AGN (e.g. Deltorn et al. 1997; 
Thimm et al. 1994).

Clusters around AGN such as MRC\,B0450--221 add to the growing number of 
distant clusters known, and offer the
same opportunities for follow-up study.
In this case, further spectroscopic observations are clearly needed 
to confirm the redshifts of other potential cluster members (including 
the red galaxies) and probe the velocity dispersion of the system,
perhaps yielding further evidence for merging.
Diffuse X-ray emission should be detectable from hot 
intracluster gas --- quasar MRC\,B0450--221 was detected in the 
ROSAT All Sky Survey  (Brinkmann, Seibert \& Boller 1994; 
Baker, Hunstead \& Brinkmann 1995) 
but deeper X-ray observations are required to disentangle the
pointlike AGN and diffuse cluster components. More detailed radio 
mapping of MRC\,B0450--221 may yield further evidence for interactions
of the radio plasma with the cluster environment.
By targetting the environments of a large sample of AGN, we will
be able to study in
detail a range of cluster environments at $z\sim 1$.

\section{Conclusions}

We have carried out a detailed study of the field of the $z=0.9$ quasar
MRC\,B0450$-$221 using narrow- and broad-band imaging and spectroscopy.
The main conclusions are:

(i) The quasar clearly lies at the centre of an overdensity of red galaxies,
with colours consistent with passively-evolving galaxies at $z=0.9$.

(ii) Tunable filter imaging finds a number of ELG
candidates with $21.0<I(AB)<22.0$ and
equivalent widths exceeding 70\AA. 
Three have been confirmed spectroscopically to lie at $z=0.9$. 

(iii) The number of emission-line objects  associated with the
quasar is considerably greater than expected from the 
average space density of star-forming field galaxies at $z=0.9$. 

(iv) The line-emitting galaxies have equivalent widths and 
inferred star-formation 
rates comparable with those of field galaxies at $z=0.9$. 

(v) The strongest line emitting galaxies, including those confirmed
spectroscopically, are located primarily in a small 
group several hundred kpc to the west of the quasar. 

(vi) Together,
these data suggest the quasar resides near the centre of a cluster 
which is accreting gas-rich galaxies from its environs. 

Observations of more $z\sim1$ quasar fields are underway to identify
and study the properties of high-redshift clusters 
and investigate their relationship with quasar activity.

\bigskip
\acknowledgements

We thank Heath Jones for many discussions.
We also thank the staff at the Anglo-Australian Telescope and at 
ESO, La Silla for their help during observing runs, and   
Pierre Leisy for his valuable assistance with the 
multiobject spectroscopy observations. 
JCB acknowledges support for this work 
which was provided by NASA through Hubble Fellowship grant 
\#HF-01103.01-98A from the Space Telescope Science Institute, 
which is operated by the Association of Universities
for Research in Astronomy, Inc., under NASA contract NAS5-26555.
RWH acknowledges funding from the Australian Research Council.

\newpage
\clearpage

%%%uncomment this line for double-spacing only.......
\renewcommand{\arraystretch}{.6}

%%
%%TABLES
%%

\begin{table}
\begin{minipage}{8.5cm}
\caption[]{Infrared Identifications ($K'>17.5$)}
\label{irtab}
\begin{tabular}{@{}rrcccccc}
\\
\tableline
\tableline
$\Delta \alpha$&$\Delta \delta$ &  
	\small	 $V$   & $I$  & $J$  & $K'$& $V-I$ & $I-K'$ \\ [2mm]
%\normalsize
\tableline
  67 &  11 & 21.8  &  20.6 &19.5 & 18.3 & 1.2 & 2.3 \\
  58 &   9 & 23.9  &  22.0 &20.6 & 18.7 & 1.9 & 3.3 \\
  55 & -10 & 21.1  &  19.6 &18.4 & 17.7 & 1.5 & 1.9\\ 
  50 &  44 & 23.0  &  21.9 &21.4 & 19.3 & 1.1 & 2.6\\
  47 &   9 &(24.8) &  22.8 &21.0 & 18.7 &(2.0)& 4.1 \\
  45 & -25 & 24.6  &  21.9 &19.7 & 18.3 & 2.7 & 3.6 \\
  45 &  53 & 22.5  &  21.4 &20.5 & 18.8 & 1.1 & 2.6 \\
  41 & -11 & 24.6  &  22.8 &20.6 & 19.1 & 1.8 & 3.7 \\
  35 & -53 & 24.2  &  21.5 &20.0 & 19.2 & 2.7 & 2.3 \\
  31 & -40 & 22.4  &  20.8 &19.8 & 18.0 & 1.6 & 2.8 \\
  30 & -42 &(24.8) &  23.2 &21.1 & 19.3 &(1.6)& 3.9 \\
  28 &   4 & 24.4  &  22.0 &19.9 & 18.2 & 2.4 & 3.8 \\
  27 & -16 &(24.8) & (23.5)&21.4 & 18.6 & 1.3 & 4.9 \\
  25 &  44 & 23.1  &  21.6 &20.1 & 18.4 & 2.1 & 3.2 \\
  24 &  -7 &(24.8) &  22.9 &21.1 & 19.0 & 1.9 & 3.9 \\
  19 &  -2 & 23.2  &  21.7 &20.4 & 18.7 & 1.5 & 3.0 \\
   8 & -59 & 23.5  &  22.0 &21.2 & 20.5 & 1.5 & 1.5 \\
   8 &  12 &(24.8) &  22.4 &21.1 & 19.6 & 2.4 & 2.8 \\
   8 &   9 &(24.8) &  23.1 &21.2 & 18.9 & 1.7 & 4.2 \\
   4 &  -5 & 24.7  &  22.7 &20.4 & 18.9 & 2.0 & 3.8 \\
   0 &   7 &(24.8) &  22.2 &20.6 & 19.3 & 2.6 & 2.9 \\ 
  -3 &  24 &(24.8) &  22.9 &20.7 & 19.0 & 1.9 & 3.9 \\
  -5 &  28 & 24.0  &  22.0 &21.3 & 20.1 & 2.0 & 1.9 \\
  -8 &  39 &(24.8) & (23.5)&21.6 & 19.5 & 1.3 &(4.0) \\
 -14 &  23 & 23.9  &  22.4 &21.6 & 19.4 & 1.5 & 3.0 \\
 -17 & -38 & 22.4  &  21.3 &20.2 & 18.7 & 1.1 & 2.6 \\
 -21 &  44 & 21.0  &  19.7 &18.9 & 18.8 & 1.3 & 0.9 \\
 -26 &  30 &(24.8) &  22.7 &20.8 & 21.5 & 2.1 & 1.2 \\
 -28 & -22 &\dots  &  21.8 &19.0 & 17.9 &\dots& 3.9  \\
 -29 &  12 &\dots  &  22.3 &21.6 & 19.2 &\dots& 3.1 \\
 -29 &  18 &\dots  &  \dots&19.3 & 17.7 &\dots&\dots \\
 -30 &   5 & 24.5  &  23.1 &21.6 & 19.2 & 1.4 & 3.9 \\
 -35 &   1 & 23.9  &  21.8 &20.2 & 18.4 & 2.1 & 3.4  \\ 
\tableline
\end{tabular}
Columns--- 
$\Delta \alpha$ (RA), $\Delta \delta$ (Dec): 
position relative to quasar (arcseconds).
\newline
$V,I,J,K'$ magnitudes: 
objects below detection threshhold (limits) 
appear in parentheses; three objects
close to bright star 
could not be measured in some bands.
\end{minipage}
\end{table}

\newpage
\clearpage

\begin{table}
\begin{minipage}{8.5cm}
\caption[]{TTF example setup parameters}
\label{tabset}
\begin{tabular}{@{}ll}
\\
\tableline

 Order-blocking filter  	&     7067/260 \AA\  \\
 Observed wavelength ($\lambda$)    &     7075 \AA\  \\
 Etalon plate spacing ($\lambda_0$) 	&     9.6 $\mu$m  \\
 Order of interference ($m$) 	&     27  \\
 Free Spectral Range ($FSR(Z)$) &     373.8 units \\
 Bandpass ($\Delta Z$) 		&     12.5  units \\
 Conversion factor (d$Z$/d$\lambda$)& 1.43  \\
 Free Spectral Range ($FSR(\lambda)$)&     261.5 \AA\  \\
 Bandpass ($\Delta\lambda$)    	&     8.7 \AA\  \\
 Effective bandpass ($\pi\Delta\lambda/2$) & 13.7 \AA\  \\
 Resolving power ($\cal R$)     &     809  \\
 Effective finesse ($N$)        &     30  \\
\tableline
\end{tabular}
Note --- parameters are given in terms of either
wavelength (\AA) or arbitrary instrumental units ($Z$)
for the Fabry-Perot etalon settings.
Standard definitions are used, as summarised for TTF by
Jones \& Bland-Hawthorn 1997, 1998 and also 
Jones, Shopbell \& Bland-Hawthorn, in preparation.

\end{minipage}
\end{table}

\newpage
\clearpage

\begin{table*}
\begin{minipage}{17cm}
\caption[]{Spectroscopic Targets}
\label{tabmos}
\begin{tabular}{@{}ccccccrcccl}
\\
\tableline
\tableline

object & TTF  & TTF  & $V-I$& TTF   & TTF  & TTF & spec &spec &spec & Comments \\ 
    & cont & line &      & sigma & line & EW  & line &EW   &$z$  & \\
    & mag  & mag  & mag  &       & (\AA)&(\AA)& (\AA)&(\AA)&     &  \\
(1) & (2) & (3) & (4) & (5) & (6) & (7) & (8) & (9) & (10) \\[2mm]
\tableline
$a$ & 21.02& 20.57& 1.2 &\llap{$<$}3&\dots&\dots&\dots&\dots&0.143&\ha, 
\nii, \sii \\
$b$ & 21.43& 20.59& 1.0 &\llap{$<$}3&\dots&\dots&\dots&\dots&0.449&\oiii, \hb \\
$c$ & 21.42& 20.56& 1.6 &\llap{$<$}3&\dots&\dots&\dots&\dots&? \\
$d$ & 21.44& 20.83& 3.0 &\llap{$<$}3&\dots&\dots&\dots&\dots&? & red galaxy \\
$e$ & 20.46& 19.97& 1.9 &\llap{$<$}3&\dots&\dots&\dots&\dots&0.663&\oiii, \hb \\
$f$ & 21.92& 20.53& 1.1 & 3 & 7084 &128& 7076  &$130\pm50$&0.899 & \oii \\
$g$ & 21.15& 20.52& 0.9 &\llap{$<$}3&\dots&\dots&\dots&\dots&? &\\
$i$ & 22.00& 20.42& 0.8 & 4 & 7077 &110& \dots &\dots&0.351 &\oiii, \hb  \\
$ii$&(22.40)&20.12& 0.7 & 5 & 7070 &(162\rlap{)}& 7066  &$70\pm25$&0.896 & \oii \\
$j$ & 21.78& 20.18& 1.1 & 5 & 7073 &163& 7086&$130\pm50$&0.901 &  \oii \\
$k$ &(22.40)&20.54& 0.9 & 3 & 7069 &(116\rlap{)}& 7057 &$30\pm10$&0.894 
& \oii ? \\
$l$ & 21.71& 20.89& 1.8 &\llap{$<$}3&\dots&\dots&\dots&\dots& ? &\\ \\
\tableline
\end{tabular}
%
%\tablenotecomments{
Columns--- (2) \& (3) Estimated magnitudes from TTF data 
for continuum level and line peak 
(objects near mag limits are shown in parentheses); 
(4) broad-band  $V-I$ colour; 
(5) significance of TTF peak detection above rms noise; 
(6) \& (7) central wavelength and observed 
equivalent width of TTF line detection 
(uncertainties 7\AA\ and about 50\% respectively; parentheses indicate
uncertain continuum mag); 
(8) \& (9) measured wavelength and equivalent width of line from spectroscopy; 
(10) spectroscopic redshift. \\
%}
\end{minipage}
\end{table*}


\newpage


\begin{thebibliography}{99}
\def\mnras{MNRAS}
\def\mn{MNRAS}
\def\apj{ApJ}
\def\aa{A\&A}
\def\nat{Nature}
\def\aj{AJ}



\bibitem[1]{} Baker J.C., Hunstead R.W., Brinkmann, W., 1995, \mnras, 277, 553

\bibitem[2]{} Balogh M.L., Morris S.L., Yee H.K.C., Carlberg R.G., Ellingson
           E., 1997, ApJ, 488, L75

\bibitem[3]{} Balogh M.L., Schade D., Morris S.L., Yee H.K.C., 
Carlberg R.G., Ellingson E., 1998, \apj, 504, L75

%\bibitem[]{} Barthel P.D., 1989, ApJ, 336, 606

%\bibitem[]{} Bertin E., Arnouts S., 1996, \aa\ Supp., 117, 393

\bibitem[1]{} Best P.N., Longair M.S., R\"ottgering H.J.A., 1998, MNRAS, 295, 549

\bibitem[1]{} Bland-Hawthorn J., Jones D.H., 1997, PASA, 15, 44

\bibitem[1]{} Bland-Hawthorn J., Jones D.H., 1998, In Optical Astronomical
Instrumentation, Kona, Hawaii, Proc. SPIE, Vol. 3355, p855

\bibitem[1]{} Bower R.G., Castander F.J., Couch W.J., Ellis R.S.,
             B\"{o}hringer H., 1997, \mn, 291, 353

\bibitem[1]{} Bremer M.N., Baker J.C., 1999, in `The Most Distant Radio
Galaxies', eds H.J.A. R\"{o}ttgering, P.N. Best, M.D. Lehnert, p425 

\bibitem[1]{} Bremer M.N., Baker J.C., Lehnert M., 2001,  in preparation

\bibitem[1]{} Brinkmann W., Seibert J., Boller Th, 1994, A\&A, 281, 355

\bibitem[1]{} Butcher H., Oemler A.Jr, 1984, ApJ, 285, 426

%\bibitem[]{} Cimatti A. et al., 1999, \aa, 352, L45

\bibitem[1]{} Clowe D., Luppino G.A., Kaiser N., Gioia I.M., 
1998, ApJ, 497, L61

%\bibitem[1]{} Colberg J.M.,  White S.D.M., Jenkins  A.,  Pearce F. R., 1999,
%MNRAS, submitted (astro-ph/9711040)

\bibitem[1]{} Connolly A.J., Szalay A.S., Dickinson M., Subbarao M.U., 
Brunner R.J., 1997, ApJL, 486, L11


\bibitem[ ]{} Cowie L.L., Hu E., Songaila A., Egami E., 1997, ApJ, 481, L9

\bibitem[ ]{} Cowie L.L., Songaila A., Barger A., 1999, AJ, 118, 603

\bibitem[ ]{} Cowie L.L., Songaila A., Hu E.M., Cohen J.G., 1996, AJ, 112, 839

\bibitem[ ]{} Deltorn J.-M., Le Fevre O., Crampton D., Dickinson M., 
1997, \apj, 483, 21

\bibitem[ ]{} Dressler A., Shectman S.A., 1988, AJ, 95, 985

\bibitem[ ]{} Dressler A., Gunn J.E., 1992, ApJS, 78, 1

\bibitem[ ]{} Eales S.A., 1985, MNRAS, 213, 899

\bibitem[ ]{} Ellingson E., Yee H.K.C., Green R.F., 1991, ApJ, 371, 49 
           %%(evoln paper)

\bibitem[ ]{} Ellis R.S., Colless M., Broadhurst T., Heyl J., Glazebrook K., 1996,
         MNRAS, 280, 235

\bibitem[ ]{} Elston R., Rieke G.H., Rieke M., 1988, ApJ, 331, L77

\bibitem[ ]{} Fioc M., Rocca-Volmerange B., 1997, A\&A, 326, 950

\bibitem[ ]{} Gallagher J.S., Bushouse H., Hunter D.A., 1989, AJ, 97, 700

\bibitem[ ]{} Gaskell C.M., 1982, \apj, 263, 79

\bibitem[ ]{} Hall P.B., Green R.F., 1998, ApJ, 507, 558

\bibitem[ ]{} Hall P.B., Green R.F., Cohen M., 1998, ApJS, 119, 1

\bibitem[ ]{} Hammer F. Flores H., Lilly S.J., Crampton D., 
Le Fevre O., Rola C., Mallen-Ornelas G., Schade D., Tresse L., 
1997, ApJ, 491, 477

\bibitem[ ]{} Henry J.P. \etal, 1997, AJ, 114, 1293

\bibitem[ ]{} Hill G.J., Lilly S.J., 1991, ApJ, 367, 1

\bibitem[ ]{} Hoessel J.G., Gunn J.E., Thuan T.X., 1980, ApJ 241, 486

\bibitem[ ]{} Hu E., Cowie L., McMahon R., 1998, ApJL, 502, L99

\bibitem[ ]{} Hu E., McMahon R., Cowie L., 1999, ApJL, 522, L9

\bibitem[ ]{} Hunstead R.W., Murdoch H.S., Shobbrook R.R., 1978, \mn, 185, 149


\bibitem[ ]{} Hutchings J.B., 1995, AJ, 109, 928 %%(z=2 cls)
 
\bibitem[ ]{} Hutchings J.B., Crampton D., Persram D., 1993, AJ, 106, 1324
           %%Galaxy clusters with QSOS at redshift 1.1 

\bibitem[ ]{} Jenkins A. et al., 1998, \apj, 499, 20

\bibitem[ ]{} Kapahi V.K., Athreya R.M., Subrahmanya C.R., Baker J.C.,
Hunstead R.W., McCarthy P.J., van Breugel W., 1998, ApJS, 118, 327 

\bibitem[ ]{} Kennicutt R.C., 1992, ApJ, 388, 310

\bibitem[ ]{} Kennicutt R.C., 1998, ARAA, 36, 189

%\bibitem[ ]{} Large M.I., Mills B.Y., Little A.G., Crawford D.F., Sutton
%      J.M., 1981, \mn, 194, 693

\bibitem[ ]{} Lilly S.J., Le Fevre O., Hammer F., Crampton D., 1996, \apj L, 
  460, L1

\bibitem[ ]{} Liu M.C., Dey A., Graham J.R., Bundy K.A., Steidel C.C., 
Adelberger K., Dickinson M.E., 2000, AJ, 119, 2556

%\bibitem[ ]{} Luppino G.A., Gioia I.M., 1995, ApJ, 445, L77

\bibitem[ ]{} Madau P., Ferguson H.C., Dickinson M.E., Giavalisco M.,
Steidel C.C., Fruchter A., 1996, MNRAS, 283, 1388

\bibitem[ ]{} Madau P., Pozzetti L., Dickinson M.E., 1998, ApJ, 498, 106

\bibitem[ ]{} Martini P., Osmer P.S., 1998, AJ, 116, 2513

%\bibitem[ ]{} McCarthy P.J., 1993, ARAA, 31, 639

\bibitem[ ]{} McCarthy P.J., Persson S.E., West S.C., 1992, ApJ, 386, 52

\bibitem[ ]{} McCarthy P.J., van Breugel W.., Kapahi V.K., 1991, ApJ, 371, 478

\bibitem[ ]{} McCracken H.J., Metcalfe N., Shanks T., Campos A., Gardner J.P., 
Fong R., 2000, MNRAS, 311, 707

\bibitem[ ]{} Morris S.L., Hutchings J.B., Carlberg R.G., Yee H.K.C.,
Ellingson E., Balogh M.L., Abraham R.G., Smecker-Hane T., 1998,
ApJ, 507, 84

%\bibitem[ ]{} Oemler A.Jr., Dressler A., Butcher H.R., 1997, ApJ, 474, 561

\bibitem[ ]{} Owen F.N., Ledlow M.J., Morrison G.E., Hill J.M., 1997, \apj,
488, L15

%\bibitem[ ]{} Pei Y.C., Fall M.S., 1995, \apj, 454, 69

\bibitem[ ]{} Pozzetti L., Mannucci F., 2000, MNRAS, 317, L17

\bibitem[ ]{} Rakos K.D., Schombert J.M., 1995, \apj, 439, 47

%\bibitem[ ]{} Rakos K.D., Maindl T.I., Schombert J.M., 1996, \apj, 466, 122

\bibitem[ ]{} Saracco P., D'Odorico S., Moorwood A., Buzzoni A., Cuby J.-G., 
Lidman C., 1999, \aa, 349, 751

\bibitem[ ]{} Schlegel D.J., Finkbeiner D.P., Davis M., 1998, \apj, 500, 525

\bibitem[ ]{} Smail I., Edge A.C., Ellis R.S., Blandford R.D., 1998,
MNRAS, 293, 144

\bibitem[ ]{} Snellen I.A.G., Bremer M.N., Schilizzi R.T., Miley G.K., 
van Ojik R., 1996, MNRAS, 279, 1294

\bibitem[ ]{} Stanford S.A., Elston R., Eisenhardt P.R., 
Spinrad H., Stern D., Dey A., 1997, AJ, 114, 2232

\bibitem[ ]{} Thimm G.J., R\"{o}ser H.-J., Hippelein H., Meisenheimer K.,
1994, \aa, 285, 785

\bibitem[ ]{} Thompson et al., 1999, \aj, 523, 100

\bibitem[ ]{} Tytler D., Fan X.-M., 1992, ApJS, 79, 1

\bibitem[ ]{} Valdes F., 1993, FOCAS User's Guide, NOAO

%\bibitem[ ]{} West M.J., 1994, MNRAS, 268, 79

\bibitem[ ]{} Yee H.K.C, Green R.F, 1987, ApJ, 319, 28

\bibitem[ ]{} Zabludoff A.I., Geller M.J., Huchra J.P., Vogeley M.S., 1993, 
       AJ, 106, 1273 

\end{thebibliography}
\end{document}